%% file: 0_ASEMain.tex
\documentclass[sigconf]{acmart}
\acmJournal{PACMPL}
\acmVolume{1}
\acmNumber{CONF} % CONF = POPL or ICFP or OOPSLA
\acmArticle{1}
\acmYear{2018}
\acmMonth{1}
\acmDOI{} % \acmDOI{10.1145/nnnnnnn.nnnnnnn}
\startPage{1}

\copyrightyear{2024}
\acmYear{2024}
\setcopyright{rightsretained}
\acmConference[ASE '24]{39th IEEE/ACM International Conference on Automated
Software Engineering }{October 27-November 1, 2024}{Sacramento, CA, USA}
\acmBooktitle{39th IEEE/ACM International Conference on Automated Software
Engineering (ASE '24), October 27-November 1, 2024, Sacramento, CA, USA}
\acmDOI{10.1145/3691620.3695473}
\acmISBN{979-8-4007-1248-7/24/10}

\usepackage{booktabs}   %% For formal tables:
\usepackage{subcaption} %% For complex figures with subfigures/subcaptions
\usepackage{graphicx}
\usepackage{textcomp}
\usepackage{xcolor}
\usepackage{url}
\usepackage{caption}
\usepackage{multirow}
\usepackage{colortbl}
\usepackage{pifont}
\usepackage{tabularx}
\usepackage{booktabs}
\usepackage{algorithm}
\usepackage{algpseudocode}
\usepackage{pifont}
\usepackage{breakcites}

\usepackage{caption}
\usepackage{subcaption}
\usepackage{amsfonts} 
\usepackage{graphicx}

\newcommand{\cmark}{\ding{51}}
\newcommand{\xmark}{\ding{55}}
\def\BibTeX{{\rm B\kern-.05em{\sc i\kern-.025em b}\kern-.08em
    T\kern-.1667em\lower.7ex\hbox{E}\kern-.125emX}}

\newcommand{\FLOW}[1]{{\color{black} #1}}

\newcommand{\revision}[1]{{\color{black} #1}}
\newcommand{\quotes}[1]{``#1''}
\newcommand{\SC}{\texttt{Scaler}}
\newcommand{\Scaler}{\texttt{Scaler}}
\newcommand{\XFA}{\texttt{XFA}}

\newcommand{\ditool}{\texttt{DITool}}

\newcommand{\perf}{\texttt{perf}}

\newcommand{\ltrace}{\texttt{ltrace}}
\newcommand{\bpftrace}{\texttt{bpftrace}}
\newcommand{\vtune}{\texttt{vtune}}
\newcommand{\vtuneums}{\texttt{vtune-ums}}
\newcommand{\parsec}{\texttt{PARSEC}}
\newcommand{\ptrace}{\texttt{ptrace}}

\newcommand{\ebpf}{\texttt{ebpf}}

\newcommand{\partialcheckmark}{\checkmark\kern-1.1ex\raisebox{.7ex}{\rotatebox[origin=c]{125}{--}}}%

\acmBadgeL[https://www.acm.org/publications/policies/artifact-review-and-badging-current]{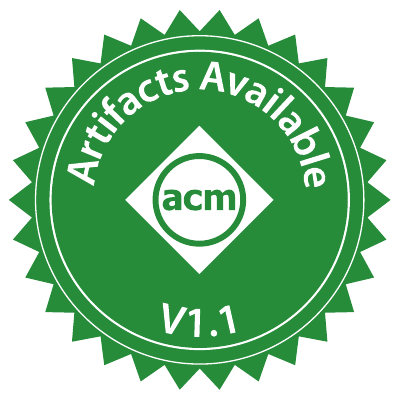}
\acmBadgeR[https://www.acm.org/publications/policies/artifact-review-and-badging-current]{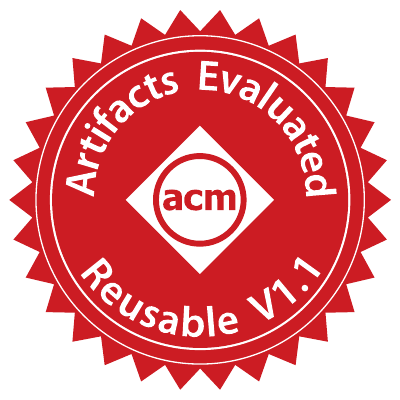}

\begin{document}

\title{Scaler: Efficient and Effective Cross Flow Analysis}

\author{Jiaxun Tang}
\email{jtang@umass.edu}
\affiliation{%
  \institution{University of Massachusetts Amherst}
  \city{Amherst}
  \state{MA}
  \country{USA}
}

\author{Mingcan Xiang}
\email{mingcanxiang@umass.edu}

\affiliation{%
  \institution{University of Massachusetts Amherst}
  \city{Amherst}
  \state{MA}
  \country{USA}
}

\author{Yang Wang}
\email{wang.7564@osu.edu}
\affiliation{%
  \institution{Meta/The Ohio State University}
  \city{Columbus}
  \state{OH}
  \country{USA}
}

\author{Bo Wu}
\email{bwu@mines.edu}
\affiliation{%
  \institution{Colorado School of Mines and HiTA AI Inc}
  \city{Golden}
  \state{CO}
  \country{USA}
}

\author{Jianjun Chen}
\email{jianjun.chen@bytedance.com}
\affiliation{%
  \institution{Bytedance}
  \city{San Jose}
  \state{CA}
  \country{USA}
}

\author{Tongping Liu}
\email{tongping.liu@bytedance.com}
\affiliation{%
  \institution{Bytedance}
  \city{San Jose}
  \state{CA}
  \country{USA}
}

\input{src/0_abstract}
\maketitle

\input{src/1_intro}
\input{src/2_BackgroundAndIdea}

\input{src/3_Implementation}
\input{src/4_evaluation}
\input{src/5_discussion}
\input{src/6_relatedwork}

\input{src/7_conclusion}

\bibliographystyle{ACM-Reference-Format}
\bibliography{src/ref/refs,src/ref/steven,src/ref/tongping}

\end{document}

%% file: src/0_abstract.tex
\begin{abstract}
Performance analysis is challenging as different components (e.g., different libraries, and applications) of a complex system can interact with each other. 
%originate from any component inside the whole software stack. 
However, few existing tools focus on understanding such interactions. To bridge this gap, we propose a novel analysis method--``Cross Flow Analysis (XFA)''-- that monitors the interactions/flows across these components. We also built the \Scaler{} profiler that provides a holistic view of the time spent on each component (e.g., library or application) and every API inside each component. This paper proposes multiple new techniques, such as Universal Shadow Table, and Relation-Aware Data Folding.
% \ST{ \textit{Lightweight PLT Interposition},  and  \textit{Multithreaded Runtime Attribution}. } 
%assembly-based
%interposition
%To reduce the performance overhead, Scaler proposes multiple mechanisms, such as the fine-grained dynamic sampling (\FGDS{}) method and online data folding. 
These techniques enable \Scaler{} to achieve low runtime overhead, low memory overhead, and high profiling accuracy. Based on our extensive experimental results, \Scaler{} detects multiple unknown performance issues inside widely-used applications, and therefore will be a useful complement to existing work. 
%(dedup: memory allocations, canneal: redblack tree, hashmap)
%without changing or recompiling the program. 
%Overall, Scaler only imposes around \todo{6.49\%} performance overhead and \todo{6.49\%} memory overhead. The experiments show that Scaler could identify multiple performance issues that cannot be diagnosed with \perf{}. 

% Overall, Scaler only imposes around 22\% performance overhead, which could identify multiple performance issues that cannot be determined using existing tools. 

\revision{The reproduction package including the source code, benchmarks, and evaluation scripts, can be found at \url{https://doi.org/10.5281/zenodo.13336658}.}

%helping users to identify performance culprits. To further reduce the interference, this pa  
%For instance, we observed that a PARSEC application -- Swaptions -- runs around $2.05\times$ slower due to its use of an inappropriate library (e.g., Hoard allocator).  
%This work aims to design transparent and automated performance diagnosis via full-stack analysis. 
\end{abstract}

%% file: src/1_intro.tex
\section{Introduction \label{sec:Introduction}}

%\ST{Reviewer suggests that we make it clear this method is based on intercepting interactions between dynamically linked libraries.}

%\ST{Reviewer suggests that we clearly define the meaning of "components".}

%\ST{Reviewer suggests that we clearly states what is "flow".}

%\ST{Scaler interception mechanism do not rely on symbol name, but users need this information to debug. So we cannot say that Scaler do not rely on symbol name. What we do not require is debugging symbol and function signature information. }

%\ST{State clearly that our goal is to profile C/C++ programs.}

%\ST{Intel PIN requires that user defines function signature. I'm wondering whether we can let the reviewer understand the difficulty caused by intercepting a function without signature. Comparing overhead with intel PIN is easy. We can just do the same thing as bpftrace.}

Modern systems are enormously complex. 
A user program typically interacts with different libraries in the whole system, such as standard libraries, the memory allocator, and other third-party libraries. Studying the interactions between these components has a two-fold indication on the performance. On the one hand, application developers may use some inappropriate library APIs with hidden performance issues. On the other hand, the extraordinary behavior of API invocations can be utilized to infer inefficient algorithm design or configurations of applications and libraries. 

\FLOW{Although a significant amount of profilers have been proposed in the past, none of them focuses on the interactions of components.} Some existing tools focus on the performance issues related to hardware, such as cache misses~\cite{Sheriff, ArrayTool, Feather, CCProf, CachePerf}; Some detect the multithreading-related performance issues~\cite{SyncPerf, wPerf, Coz}, mainly on thread-related APIs; Some may report the time spent on user functions via the sampling mechanism~\cite{OProfile, gprof, vtune-ums}. \ltrace{} is possibly the most related work~\cite{ltrace}, but introduces over $6195\times$ performance overhead. Such high overhead makes it implausible to identify issues caused by library APIs accurately.  

\begin{table}[!th]
\setlength{\tabcolsep}{4pt}
\scalebox{0.75}{
\begin{tabular}{llccc}
    \hline
    \multirow{2}{*}{Instrumentation Technique} & \multirow{2}{*}{Tool} & \multirow{2}{*}{\begin{tabular}[c]{@{}c@{}}Collection\\ Frequency\end{tabular}} & \multicolumn{2}{c}{Overhead} \\ \cline{4-5} 
     &  &  & Runtime & Memory \\ \hline
    Sampling via Hardware PMUs & \perf{} & $0.2\%$ & $22.5\%$ & $6.7\times$ \\
    Sampling via timer + \ptrace{} & \vtuneums{} & $0.0\%$ & $41.4\%$ & $18.3\times$ \\
    \texttt{eBPF} + software breakpoint & \bpftrace{} & $100\%$ & $28.8\times$ & $3.1\times$ \\
    \ptrace{} + software breakpoint & \ltrace{} & $100\%$ & $>6195.7\times$ & - \\ \hline
    Universal Shadow Table & \Scaler{} & $100\%$ & $20.3\%$ & $15.5\%$ \\ \hline
    \end{tabular}
}
\caption{Compare with existing work. \label{tbl:BaselineComparison}}
\end{table}

%As the effect, \FLOW{none of existing is challenging yet essential to identify the performance issues caused by the cross-component interactions in the whole system stack.}

We propose a novel method called \textbf{\textit{Cross Flow Analysis} (\XFA{})} that monitors and analyzes the interactions  between different components (called \textbf{cross flow})  in the system, where the component is either the application itself or any library. \XFA{} helps identify potential performance problems caused by libraries (e.g., incorrect API uses or inappropriate configuration), and inefficient algorithm design of applications (e.g., data structures). More specifically, \XFA{} proposes to trace all Application Interfaces (APIs) of involved libraries, including the number and runtime of invocations of each API. To assist the performance analysis, \XFA{} summarizes the runtime/invocations of each API and each component and provides two views: a component view shows \textit{the time (and percentage) of one component spending on other related components}, and an API view of a component shows \textit{the time (and percentage) spent on any API inside this component.} 

\begin{figure}[!tbp]
\centering
\includegraphics[width=2.5in]{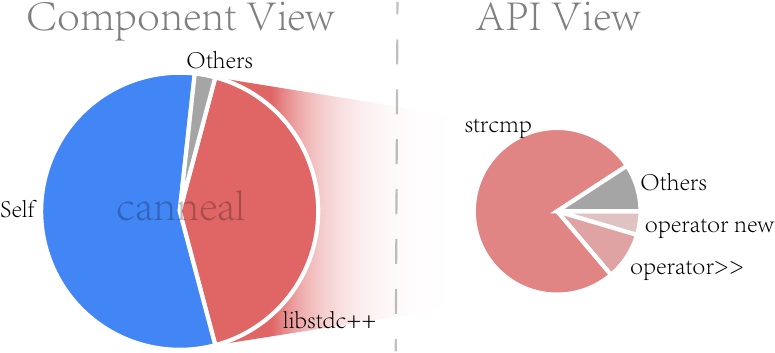}
\caption{\Scaler{}'s report for  \texttt{canneal}. }
%{\color{red}Yang: Similarly, one who is familiar with performance analysis to some extent may feel surprised that existing tools like perf cannot provide such information. So if perf has a high overhead/less accuracy/requires recompilation, etc, it may be worthwhile emphasizing its drawback here.}
\label{fig:intro_xfa_basicidea_b} 
\Description[\Scaler{}'s report for  \texttt{canneal}.]{\Scaler{}'s report for  \texttt{canneal}.}
%\todo{needs to change}
\end{figure}

Figure~\ref{fig:intro_xfa_basicidea_b} shows an example report for the \texttt{canneal} application of PARSEC~\cite{bienia2008parsec}, which is a \textbf{new} bug reported by \SC{}: the \texttt{libstdc++} library is the most time-consuming library (from the component view), 
%where \texttt{canneal} spends $25\%$ of time on the \texttt{libstdc++} library. 
and \texttt{strcmp} consumes $77\%$ runtime of \texttt{libstdc++} (from the API view). Since \texttt{canneal} simulates an algorithm that optimizes the routing cost of circuit boards, it is certainly \textit{abnormal} to have such a large portion of time spent on string comparisons. \textit{Such an issue originates from an inappropriate design} when \texttt{canneal} counts the appearance of each string in the input file: the std::map (red-black tree) is used to store the strings, where the searching/inserting operations and tree balancing operations incur a significant number of string comparisons; Instead, a better method is to utilize \texttt{std::unordered\_map} (hash map). Changing the data structure not only reduces string compares, but also reduces the last-level cache misses by $18\%$. The combined effect improves the performance by $52\%$. That is, the abnormal behavior of library APIs helps expose the inefficient algorithm design inside user programs. In contrast, \perf{} fails to detect this issue because of its excessively coarse sampling rate, resulting in the inaccurate report of the runtime for \texttt{libstdc++} and \texttt{strcmp}. It reports that only 6.32\% of time is spent in \texttt{libstdc++} and only 0.56\% time is spent in \texttt{strcmp}. 
%, possibly due to its too-coarse sampling and incorrect handling of multithreaded events.  

%\texttt{libhoard} (a memory allocator library) is the performance culprit, since 40\% of execution time of the application is spent on this library. By checking the component view of \texttt{libhoard}, as shown in the top-right pie chart, it is clear that \texttt{libhoard} spends 47\% of its time on the \texttt{libpthread} library, which is very abnormal. With the  API view (bottom-right pie chart), it is easy to identify the final root cause, which is caused by the lock contention. 

%\TP{Should we use the example of finding inherent algorithm design instead, which is more impressive and none of existing tools could identify such an issue.}

Based on the idea of \XFA{}, we further built a profiler -- \Scaler{} -- that monitors API invocations. Our profiler requires overcoming the following technical challenges. 
%efficiently. To make it practical for profiling interactions between arbitrary components across the software stack, \Scaler{} requires no change or recompilation of applications and libraries. Overall, \Scaler{} requires to overcome multiple technical challenges. 

%\ST{Reviewers did not get that tools like gdb only use ptrace for several APIs so they can be fast. We must find some ways to emphasize that our purpose is to provide global view.}

\textit{Challenge 1: how to trace different types of library API invocations without recompilation and knowing the signatures of APIs?}  
%Existing tracing techniques, such as compiler instrumentation, general binary instrumentation, process-based tracing, and linkage interposition, as discussed in Section~\ref{sec:RelatedWork}, have their own issues. The 
Compiler-based instrumentation requires to recompile all involved components~\cite{gprof}, which is often infeasible because of the lack of source code, build scripts and proper build environments. General binary instrumentation (e.g., Pin~\cite{Pin}) and \ptrace{}-based technique (e.g., \texttt{ltrace}{}~\cite{ltrace}) easily introduce orders of magnitude performance overhead. \ditool{}~\cite{ditools} requires signatures of the profiled APIs to hook APIs. 
This paper proposes a light-weight binary instrumentation technique to trace different types of APIs. For dynamic linking, \Scaler{} replaces binary entries of the related ELF sections in memory; For dynamic loading, \Scaler{} intercepts \texttt{dlsym} (and \texttt{dlopen}) so that it can redirect API invocations to the universal interceptor, as described in Section~\ref{sec:BasicIdeaOfSC}.  
%\Scaler{} intercepts , including dynamic linking via Procedure Linkage Table (PLT) and Global Offset Table (GOT)~\cite{pltgotdoc}, and dynamic loading via \texttt{dlsym}~\cite{dlsym}. Dynamic linking further includes eager and lazy binding~\cite{lazybinding}. 
% \texttt{.got.plt} and \texttt{dlsym} (and \texttt{dlopen}) are returning the real PLT 
% \ST{The very last sentence seems confusing to me.}
%the entries of \texttt{.plt} and \texttt{.got.plt} table in ELF files  }

%\Scaler{} intercepts , including dynamic linking via Procedure Linkage Table (PLT) and Global Offset Table (GOT)~\cite{pltgotdoc}, and dynamic loading via \texttt{dlsym}~\cite{dlsym}. Dynamic linking further includes eager and lazy binding~\cite{lazybinding}. 

\begin{figure}[ht]
\centering
\includegraphics[width=3.2in]{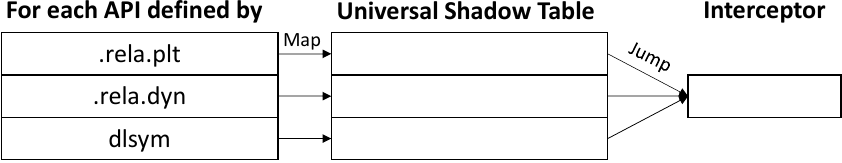}
% \captionsetup{justification=justified}
\caption{Overview of Universal Shadow Table. \label{fig:intro_UniversalShadowTable}}
\Description[Overview of Universal Shadow Table.]{Overview of Universal Shadow Table.}
% \todo{show the relationship of ``universal'' by using the PLT table, got.plt entry, and dlsym function.  }
\end{figure}

\textit{Challenge 2: how to trace API invocations with low runtime overhead?} 
\Scaler{} utilizes the same interceptor to handle different types of API invocations, but requires storing the events of each API invocation separately and keeping the invocation hierarchy. To satisfy this requirement,  \textbf{Universal Shadow Table} is proposed to efficiently intercept APIs, as shown in Figure~\ref{fig:intro_UniversalShadowTable}. For each API defined by \texttt{.rela.plt}, \texttt{.rela.dyn} and dynamically loaded by \texttt{dlsym}, \Scaler{} maps them to a shadow entry in the Universal Shadow Table. Each shadow entry consists of multiple binary instructions that can store API-specific information required by \Scaler{}. By using Universal Shadow Table, \Scaler{} only introduces around $20\%$ overhead but intercepts around $63$ million invocations per second. Key design choices will further be discussed in Section~\ref{sec:BasicIdeaOfSC}.

%via two instructions (less than 16 bytes), and then utilizes a shared interceptor that can handle all APIs through the same code. 
%, since the original entry (usually 16 bytes) is insufficient to enclose all necessary information for the interception,  Instead

% .got.plt: lazy and eager
% .plt.got: eager loading
% \ST{Since .plt.got is not mentioned in the text, do we still draw ".plt.got" in the graph? Or should we find a better way to distinguish ".plt.got" and ".got.plt" in the text?}

\textit{Challenge-3: how to record intensive API invocations with minimal storage overhead while maintaining accuracy?} One common method of tracing used by \ltrace{}~\cite{ltrace} is to append one event after the other. However, recording all API invocations will impose high storage and performance overhead. This overhead is caused by frequent API invocations at around $63$ million per second. 
Instead, \Scaler{} proposes the \textbf{Relation-Aware Data Folding} with the following attributes: (1) During the recording phase, all API invocations will be summarized at runtime and output at the end. This mechanism prevents a proportional increase in storage/memory volume over time. (2) The recording will maintain the relative relationship between APIs and libraries. In cases where the same API is invoked by different libraries, the summary will only group invocations originating from the same library together. This attribute helps identify the performance problem inside one specific component, that is, maintaining the accuracy. Overall, \Scaler{} only imposes around $16\%$ memory overhead via its Relation-Aware Data Folding.

%\Scaler{} cannot rely on debugging symbol or source code information to perform API interception because third-party libraries often does not have them. To work correctly, \Scaler{} needs to seamlessly integrate into the API invocation process and handle many corner cases imposed by existing dynamic linking mechanisms. Key methods will further be discussed in Section \ref{sec:DesignAndImpl_ReliableDataRecording}. 

%\textit{challenge-3} let the reviewer understand the difficulty caused by intercepting a function without signature. How to perform the interceptions on any apis in general? 

%\textit{Challenge-3: How to model API invocations of multithreaded applications?} It is evident that the same amount of slowdown in an API will have different effects on the overall performance depending on the number of threads running in parallel. However, considering the impact of threads will unavoidably introduce shared variable access and require proper thread synchronization. Otherwise, data race will occur and harm recording accuracy. Such synchronization must be implemented both correctly and very efficiently to avoid significant slowdown. \Scaler{} proposes \textit{\textbf{Multithreaded Runtime Attribution}} method to precisely reflect the impact of multi-threaded parallelism on time while only increasing the overall runtime overhead by 5\%. Key design choices will further be discussed in Section \ref{sec:DesignAndImpl_MultiThreadedTimeAttribution}. 

\Scaler{} also includes other contributions, such as handling the runtime attribution of the multithreaded applications, and handling corner cases like irregular API invocation. Based on our extensive evaluations, \Scaler{} imposes around $20\%$ performance overhead. \Scaler{}'s memory overhead is around $16\%$, which is orders of magnitude less than \perf{}, where \perf{} imposes around $7\times$ memory overhead. Overall, \Scaler{} identified six bugs in the evaluated applications, whereas \perf{} could only find one of them. %{\color{red} Yang: is this because perf sampling does not record all? Would they be captured if we run perf for longer period of time?}
Among these bugs, two were unknown bugs. \FLOW{Overall, this paper makes the following contributions:}
\begin{enumerate}
    \item It proposes a novel \textbf{Cross-Flow Analysis (\XFA{})} me\-thod to help users understand cross-component interactions. Such cross-flow analysis will benefit the identification of inappropriate API usage, and help identify some incorrect algorithm designs and configurations.  
    
    \item It proposes a series of novel techniques together to address the performance and memory challenges, including but not limited to \textbf{Universal Shadow Table}, and \textbf{Relation-Aware Data Folding}.  
    %It proposes a series of novel instrumentation,  data collection, and data processing methods together to address the performance and memory challenges of the full trace of API invocations, including but no limited to \textbf{Selective Binary Instrumentation}, \textbf{Universal Shadow Table}, and \textbf{Relation-Aware Data Folding}.  
    %proposes a series of novel data collection and processing methods to collectively address the performance and memory challenges while effectively profiles cross-component interactions without losing ease-of-use and extensibility. The proposed methods mainly include \textbf{\textit{Lightweight PLT Interposition}}, and \textit{\textbf{Multithreaded Runtime Attribution}}. 
  
    \item Extensive experiments have been performed on a range of real-world applications. These experiments show that \Scaler{} imposes low performance and memory overhead while effectively detecting multiple unknown performance issues related to cross-component interactions. 
    
\end{enumerate}

\FLOW{The remainder of this paper is organized as follows.} Section~\ref{sec:BkgAndOverview} provides an overview of our approach. Section~\ref{sec:DesignAndImpl} further discusses implementation details. Section~\ref{sec:Evaluation} presents our effort to evaluate the effectiveness and runtime/memory overhead of \Scaler{}. Section \ref{sec:Discussion} discusses the compatibility, extensibility, and limitations of \Scaler{}. Finally, Section~\ref{sec:RelatedWork} reviews related work, and Section~\ref{sec:Conclusion} concludes the paper.

%% file: src/2_BackgroundAndIdea.tex
\section{Background and Overview}\label{sec:BkgAndOverview}

%\ST{Add a dedicated section to explain the output format.}
In this section, we briefly discuss the background of API invocation mechanisms. Then, we introduce the basic idea of \XFA{} and \SC{}. 

\subsection{Background of API Invocations}
\label{sec:background}

In current software systems, programmers are not required to write all programs from scratch. Instead, they could employ external libraries to quickly develop the software system by invoking ``Application Programming Interface (API)''.  APIs define the methods and data formats that applications can use to request and exchange information with external components or libraries. There are two common mechanisms for invoking external APIs: dynamic linking and dynamic loading, as detailed in the following.

\subsubsection{Dynamic Linking}

In dynamic linking, libraries are not included directly in the executable binary during compilation (unlike static linking). Instead, the references to the specific functions or symbols in external libraries are resolved at runtime. 
It typically relies on a dynamic linker (\texttt{ld-linux.so}), often referred to ``linker'' interchangeably in this paper, to resolve symbol addresses in executable files and shared libraries during program execution. Dynamic linking related APIs can be categorized by two sections in the ELF file: \texttt{.rela.plt} and \texttt{.rela.dyn}.

For APIs defined by the \texttt{.rela.plt}, there are two address resolution modes~\cite{levine2001linkers}: in eager mode, the dynamic linker resolves the address of all APIs before the program execution. In lazy mode (the default mode), the dynamic linker will postpone the address resolution of an API until the first invocation time. Every time a component (the application or a library) invokes an API, the linker will first determine whether the memory address of that API has been resolved. If not, it will resolve the address with the help of ELF section \texttt{.plt} and \texttt{.got.plt}. 

\begin{figure}[!h]
\centering
\includegraphics[width=3.2in]{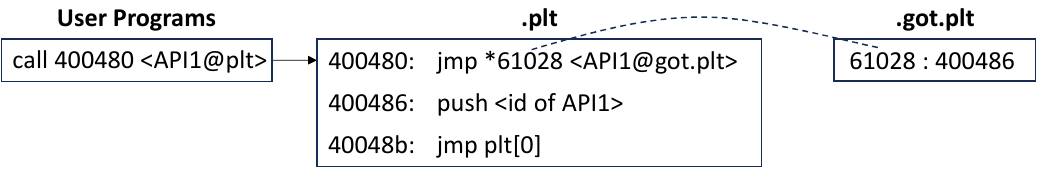}
\caption{\texttt{.rela.plt} API Invocation via \texttt{.plt} and \texttt{.got.plt}.}
\label{fig:relaPLTMeahcnism}
\Description[\texttt{.rela.plt} API Invocation via \texttt{.plt} and \texttt{.got.plt}.]{\texttt{.rela.plt} API Invocation via \texttt{.plt} and \texttt{.got.plt}.}
\end{figure}

 The whole process is illustrated in Figure~\ref{fig:relaPLTMeahcnism}. Each \texttt{.plt} entry has three executable instructions, while each entry in \texttt{.got.plt} only stores one address.
In particular, each \texttt{.plt} entry will have the following three instructions: the first instruction is a jump instruction (like ``\texttt{jmp *0x61028}'') that fetches the memory address stored in the \texttt{.got.plt}. The other two instructions will push the index of the \texttt{.plt} entry onto the stack (e.g.,  ``\texttt{push x}'') and then jump to the starting entry of the \texttt{.plt}, which is used to invoke the linker. The linker will replace the \texttt{.got.plt} entry with the resolved address before jumping to the API, so subsequent API invocations will only need to execute the first jump instruction in the \texttt{.plt} entry. For eager mode, the \texttt{.got.plt} entry will be filled with the real API address before the main function starts. In this way, the jump instruction (like ``\texttt{jmp *0x61028}'') will directly invoke the API's address.
%Overall, instructions in each PLT entry occupy 16 bytes. 

\begin{figure}[!h]
\centering
\includegraphics[width=3in]{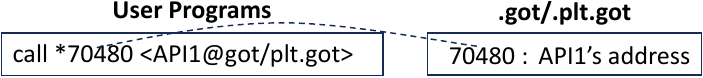}
\caption{\texttt{.rela.dyn} API Invocation via \texttt{.got} and \texttt{.plt.got}.}
\label{fig:relaDynMechanism}
\Description[\texttt{.rela.dyn} API Invocation via \texttt{.got} and \texttt{.plt.got}.]{\texttt{.rela.dyn} API Invocation via \texttt{.got} and \texttt{.plt.got}.}
\end{figure}

For APIs defined by the \texttt{.rela.dyn}, the linker will always resolve the API address before the main function starts. But the linker will store resolved address inside \texttt{.plt.got} or \texttt{.got} rather than the ELF sections corresponding to the \texttt{.rela.plt}. The instructions used to call APIs defined by \texttt{.rela.dyn} are also different. The whole process is illustrated in Figure~\ref{fig:relaDynMechanism}. The call instruction will fetch the address stored in \texttt{.plt.got} or \texttt{.got} and directly jump to it. Another prominent difference between \texttt{.rela.dyn} and \texttt{.rela.plt} is that \texttt{.rela.dyn} not only defines API functions but also global variables, while \texttt{.rela.plt} only defines API functions.

\subsubsection{Dynamic Loading}

Dynamic loading is the process of loading a shared library into memory at runtime after the program has started executing. In particular, the user program first invokes \texttt{dlopen} to get the handle of a specific library (by passing the name of the shared library file), and then explicitly requests the address of a function by invoking the \texttt{dlsym} function.

\subsection{Cross-Flow Analysis (\XFA{})}
\label{sec:BasicIdeaOfXFA}

As discussed above, \XFA{} monitors interactions between different components in the whole system stack, where each component (either the application or a library) is treated as an island. We observe that API invocation is the way for a component to interact with the external world. Therefore, we propose to intercept all API invocations in order to collect the data for performance analysis. More specifically, \XFA{} collects the accurate runtime and number of invocations for each API. In the end, \XFA{} provides two views to help the user diagnose the performance issue. One is \textit{component view}, which provides the following information: how much time one component spends on itself and other components. Another is \textit{API view} that presents the runtime distribution information of all APIs in a library.

\subsection{Overview of \Scaler{}} 
\label{sec:BasicIdeaOfSC}

\SC{} is a detailed implementation/profiler of \XFA{}. \SC{} aims to satisfy the following requirements:

\begin{itemize}
    \item \textbf{Least Manual Effort:} To use \SC{}, users do not need to recompile or change any code or use custom OS or hardware.
    
    \item \textbf{Maximum Generality:} \SC{} should support different types of API invocations, such as dynamic linking and loading, but does not rely on customized linker or the availability of symbol information or the source code. 
    
    \item \textbf{Low Recording Overhead:} \SC{} should not introduce high performance and memory overhead that can prevent its usage in the production environment, even given a significant number of API invocations.
    
\end{itemize}

In order to achieve the least manual effort, we could not employ the compilation-based approach as that requires the source code, build scripts, and the proper system environment,  which is not always accessible. We also cannot utilize the preloading technique, as that will require \Scaler{} to know all function signatures beforehand. To satisfy maximum generality, we should not rely on a particular version of dynamic linker/loader or hardware performance counters. Further, existing hardware performance counters typically utilize sampling as the basic method, which cannot provide the full tracing functionality.  After excluding all profiling methods, only binary instrumentation (like Pin~\cite{Pin} or DynamoRIO~\cite{DynamoRIO}) and software-based process tracing (e.g., \texttt{ptrace}~\cite{ptrace}) exist in existing work. However, both general binary instrumentation (like Pin~\cite{Pin} or DynamoRIO~\cite{DynamoRIO}) and process tracing are notoriously known for their high performance overhead, easily over $20\times$, which is even prohibitively high for development phases. 

To meet all of these requirements, \SC{} employs a Selective Binary Instrumentation strategy, that only instruments the locations related to API invocations. By using selective binary instrumentation, there is no need to recompile the code, rely on the custom hardware, and couple with a specific version of the linker or loader.  By limiting the scope of instrumentation, \SC{} reduces the instrumentation overhead compared to general binary instrumentation that instruments or checks the whole binary. Further, \SC{}'s selective binary instrumentation supports different modes of dynamic linking and dynamic loading. For APIs defined by \texttt{.rela.plt}, \SC{} instruments the binary of the corresponding \texttt{.plt} entries. For APIs defined by \texttt{.rela.dyn}, the address resolution occurs before the program's execution begins. For these scenarios, \SC{} performs the binary instrumentation by changing the corresponding \texttt{.got}/\texttt{.plt.got} entry directly. \SC{} instruments within the program's startup code, just before the \texttt{main()} function, so that the address resolution of these APIs has been completed. For dynamic loading, since \texttt{dlsym()} typically returns the resolved address to the program, \SC{} re-direct the return addresses to the custom common interceptor by instrumenting and intercepting the \texttt{dlsym} invocations.   

\textit{Note that selective binary instrumentation alone is not sufficient to guarantee the low runtime overhead}. In addition to the instrumentation, \SC{} requires handling each API differently, such as storing the events of each API invocation, and then returning back to the invocation site. Unfortunately, it is not easy to complete these tasks efficiently. For instance, existing library interposition approach, like \ditool{}~\cite{ditools}, redirects \texttt{.got.plt} entries to custom functions directly. \ditool{} requires the user to redefine each custom function with the same signatures so that they can handle the above-mentioned tasks, which is clearly not generalizable enough. In the development of \SC{}, we have tried to utilize the hash table to locate the location of storing events, but this method imposed a large overhead when multiple entries are mapped to the same bucket. 

To overcome the generalization and performance issue, \SC{} proposes \textbf{Universal Shadow Table} that maps each API of different types to one shadow entry in the Universal Shadow Table, as shown in Figure \ref{fig:intro_UniversalShadowTable}. The shadow entry encloses all necessary information for each specific API, such as the storing location, the jumping target, and the returning target after the interception. That is, \SC{} only needs to rely on the shadow entry to parse the callee (API) information in constant time. 
Since the size of each shadow entry is much larger than the size of a normal entry (e.g., 16 bytes for a \texttt{.plt} entry), it can include more information inside, overcoming the size limit of the original entry. Universal Shadow Table also makes it possible to utilize a common interceptor (as shown in Figure~\ref{fig:intro_UniversalShadowTable}) to handle different types of APIs. Overall, the Universal Shadow Table achieves the maximum generality and low performance overhead.

Another potential issue for the profiling is the memory or storage overhead, as existing work typically appends the recorded events one after the other, causing a proportional increase in storage/memory volume over time. \SC{} proposes the \textbf{Relation-Aware Data Folding} to reduce its memory/storage overhead while maintaining its accuracy. This design is based on the observation of \SC{}'s major purpose: \SC{} requires the understanding of the time (and percentage) one component spent on other related components (component view), and the time (and percentage) spent on a specific API inside a library (API view). Therefore, \SC{} could summarize the events of each API together, 
instead of appending the events into the log file and performing the analysis offline. By summarizing the events of each API together, \SC{}'s storage overhead does not increase proportionally over the total runtime. As mentioned above, \SC{}'s recording maintains the relative relationship between APIs and libraries. In cases where the same API is invoked by different libraries, the summarization will only group invocations originating from the same library together. That is the reason why such a method is called `Relation-Aware Data Folding''.

%% file: src/3_Implementation.tex
\begin{figure*}[ht!]
% \centering
\scalebox{0.70}{
\includegraphics[width=6.5in]{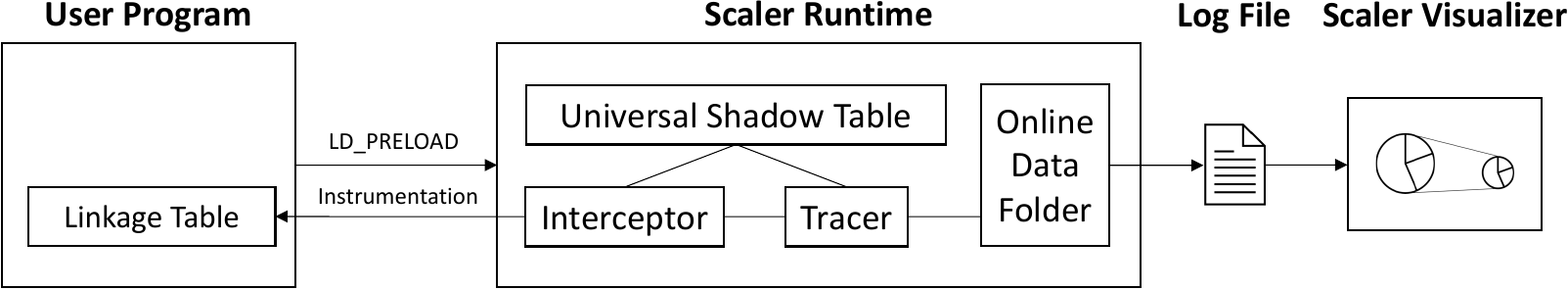}
}
\caption{Overview of \Scaler{}. \label{fig:MainComponentsOfScaler} }
\Description[Overview of \Scaler{}.]{Overview of \Scaler{}.}
\end{figure*}

\section{Design and Implementation \label{sec:DesignAndImpl}}

In this section, we introduce the major components and the implementation of \Scaler{}.

As shown in Figure~\ref{fig:MainComponentsOfScaler}, \Scaler{} includes a runtime library and an offline visualizer. To use \Scaler{}, there is no need to recompile and change user programs and any library. \Scaler{} can be linked with the user program by specifying \Scaler{}'s executable via the \texttt{LD\_PRELOAD} environment variable.  The runtime library further includes multiple components:  \texttt{Interceptor}, \texttt{Tracer}, \texttt{Universal Shadow Table}, and \texttt{Online Data Folder}. These components interact as follows: The interceptor handles different modes of API invocation and redirects the execution to the \texttt{Universal Shadow Table}. The \texttt{Universal Shadow Table} contains assembly code that can record the necessary information needed by the \texttt{Tracer}. The \texttt{Tracer} is responsible for tracing and collecting the information of each API invocation. The \texttt{Online Data Folder} is responsible for storing the events in a memory-efficient way and outputs all events to an external log file at the end of the execution. 
\Scaler{} also includes an offline visualizer to analyze the data and generate component and API views offline. As \Scaler{}'s online data folder already summarizes the recorded data online, the visualizer can analyze the recorded data very quickly. 

\subsection{Interceptor}
\label{sec:interceptor}

As mentioned above, \Scaler{} intercepts different types of API invocations caused by dynamic linking and dynamic loading. The purpose of the interceptor is to redirect different types of API invocation to the \texttt{Universal Shadow Table}.

\subsubsection{Dynamic Linking}

In UNIX-like systems, dynamic linking defines API in two areas: \texttt{.rela.plt} and \texttt{.rela.dyn} as detailed in Section~\ref{sec:background}. 
% \ST{It is maybe confusing to call .got.plt using its full name?}

For APIs defined by the \texttt{.rela.plt} section, there are two different modes: eager and lazy mode. In eager mode, the APIs' addresses have been resolved before entering the main routine, whereas in lazy mode, the address resolution will not happen until the program reaches the \texttt{.plt} entry. Despite this difference, the user programs always use the same instruction (e.g., ``\texttt{call func@plt}'') to invoke APIs. That is, both modes will execute instructions in the \texttt{.plt} section. The \texttt{Interceptor} intercepts these APIs by replacing each entry of \texttt{.plt} with two instructions, which will redirect the execution to the \texttt{Universal Shadow Table}. 

%In lazy mode, the memory address of an API function is not resolved. \Scaler{} lets the dynamic linker resolve that address, as discussed in Section~\ref{sec:shadowtable}. 

%\TP{How to resolve the address? Should we wait for the results of the linker, or we will perform the linker's resolution.} 

%\ST{
%\Scaler{} resolves address by book-keeping the second (push <id of API>) and the third (jmp PLT[0]) instructions and simulates these two instructions in the shadow PLT table. For example, to resolve the address of an API, we simply need to push the id into the stack and jump to PLT[0] of that library, and ld.so will resolve the address and put it into the .got entry}

For APIs defined by the \texttt{.rela.dyn} section, the user program will use a different instruction (e.g., ``\texttt{call\- *\-func\-@\-plt\-.got}'') to invoke the API. That is, the API invocations will bypass the \texttt{.plt}. The linker will always resolve the APIs' address before the \texttt{main} routine and place them in \texttt{.plt.got}/\texttt{.got}. To intercept APIs defined by the \texttt{.rela.dyn} section, \Scaler{} directly replaces the content in \texttt{.plt.g\-ot}/\texttt{.got} with the address of the \texttt{Universal Shadow Table}.                                                          

\subsubsection{Dynamic Loading}

API invocations of dynamic loading typically go through \texttt{dlopen} and \texttt{dlsym}. These two functions are intercepted by instrumenting the corresponding \texttt{.plt} entries as discussed above. 

%\TP{should we handle differently for dlopen? If we don't need to check the correctness, then there is no need to do that. That is, we don't need to use the custom function for dlopen.} 

\Scaler{} defines its custom \texttt{dlopen} function. When \texttt{dl\-open} is invoked, \Scaler{} not only checks the just opened library but also the dependency libraries imported implicitly by the new library. \Scaler{} will scan all newly imported libraries and hook their APIs as well.

%\ST{dlopen should be handled differently because we need to detect whether there are other dependency libraries introduced implicitly by dlopen. Because we cannot tell the dependency libraries of the dlopened library from dlopen parameters. And we need to scan these implicitly imported libraries and hook their APIs. }

\Scaler{} defines its custom \texttt{dlsym} function. When \texttt{dl\-sym} is invoked, \Scaler{} allocates an entry in the \texttt{Universal\- Shadow\- Table} and then returns the entry's address to the user programs. That is, whenever user programs invoke the function pointer returned by \texttt{dlsym}, the control logic will be passed to the \texttt{Universal Shadow Table}. Note that for APIs invoked by \texttt{dlsym}, there is no way to know beforehand which APIs will be opened this way. Therefore, \Scaler{} can only allocate shadow entries on demand. 

\subsubsection{Handling Abnormal Cases \label{sec:HandleAbornomalCase}}

In order to ensure high reliability, \Scaler{} also handles the following abnormal cases.  

\paragraph{Supporting irregular API invocation:} Some compiler optimizations may cause the program to use \texttt{jmp API@plt} instruction rather than \texttt{call API@plt} instruction when invoking APIs. With \texttt{jmp} instruction, the return addresses will not be pushed to the stack and the API will never return. \Scaler{} detects this problem by comparing the return address location on the stack. \texttt{call} instruction will always push the address of the next instruction as the return address to the stack, so if we observe that two consecutive API invocations has the same return address stored at the same location on the stack, then \Scaler{} will know the API is invoked with \texttt{jmp} instruction and will not return as well.

% user programs
%   1: call API-A
%    :  next;  <-- rsp 

%   2: In API-A, jump API-B 
  
% common handler: 
%     prehook: if the return address of current 
% API is the same the previous one, 
%     api
%     posthook:

%\TP{I totally don't understand this solution. \Scaler{} detects this problem by comparing the return address. If the return addresses for two subsequent APIs are the same, then \Scaler{} only returns once and records that the two API finishes simultaneously. }

%\ST{If a program uses "jmp" to invoke an API, then next instruction after jmp will not be pushed to the stack as return address. So, the jmp instruction will have the same return address as the previous layer of callstack. We can detect this issue by comparing whether the current return address is the same as previous layer of API call. If the return address is the same, then it means there are jump instructions and we only need to return once.}

\paragraph{\FLOW{Support no-return APIs:}} Some \texttt{glibc} APIs never return back to the caller. One typical no-return API is \texttt{exit()}, which terminates the program or a thread and will not cause performance overhead. \Scaler{} chooses not to intercept all functions marked by \quotes{\_\_noreturn} in the \texttt{glibc} library, because like \texttt{exit}, no-return APIs in \texttt{glibc} will not be the root cause of performance problems and often does not work like usual functions. If the user program invokes a no-return API, \Scaler{} can still work correctly by comparing the return location as mentioned above. 

\paragraph{\FLOW{Identify functions in \-.re\-la\-.dy\-n:}} As mentioned in Section~\ref{sec:BkgAndOverview}, \texttt{.re\-la\-.dyn} not only defines API functions but also global variables. And there is no flag to effectively distinguish the two. \Scaler{} detects whether an entry defined by \texttt{.rela.dyn} is API by checking whether it points to an executable memory region.

\subsection{Universal Shadow Table (UST)}
\label{sec:shadowtable}

The Universal Shadow Table holds one shadow entry for each API defined in \texttt{.rela.plt}, \texttt{.rela.dyn} and  dynamically loaded via \texttt{dlsym()}, as shown in Figure~\ref{fig:intro_UniversalShadowTable}. The UST is critical in \Scaler{}'s runtime.   
% \TP{Initially, the interceptor allocates a shadow entry in the table, and links every entry (as mentioned above) with the shadow entry via selective binary instrumentation and \texttt{dlsym} interception. During the invocation of every API, the execution will be transferred to the tracer based on the shadow entry. The tracer updates and folds the data according to the configuration of the UST entry. }

%\Scaler{} will generate a \texttt{push} instruction and use the unique API identifier as its immediate operand. 

Each shadow entry includes a set of assembly instructions (134 bytes in total) that implements the following functionalities, before jumping to the \texttt{Tracer}. 

\begin{itemize} 
\item \textbf{Reading per-thread context in TLS ($20$ bytes):} \Scaler{} checks the per-thread context at first. If the context is not initialized, 
%\ST{which will only occur once right before a thread begins to execute}, 
\Scaler{} skips the tracing and invokes the corresponding API directly. Note that \Scaler{} keeps the per-thread data for all API invocations. If the per-thread context is not initialized, then it is infeasible to update the tracing data. 

\item \textbf{Increment the number of API invocations and record timestamp ($45$ bytes):} \Scaler{} increments the number of invocations for the current API. By default, \Scaler{} always records the number of invocations. At the end of this code section, \Scaler{} checks whether it is necessary to perform the timing, if yes then \texttt{UST} will invoke \texttt{Tracer} to record the timestamp before API execution. For extensibility, \Scaler{} allows users to configure the frequency of collecting the runtime. 

\item \textbf{Invoking the real API ($31$ bytes):} \Scaler{} only invokes the real API directly when the per-thread context is not initialized or timing is not required based on the setting. \Scaler{} handles differently for different types of API invocations. For dynamic loading (via \texttt{dlsym}) or after address resolution in dynamic linking, the execution will jump to the API address directly. For API invocation in dynamic linking (before address resolution), the \texttt{UST} will simulate the behavior of the original \texttt{.plt} entries and invoke \texttt{ld-linux.so} to resolve the address. The return address of the real API is recorded and replaced with the current \texttt{UST} entry so that the real API will return to the \texttt{UST} after execution. Note that before invoking the real API, the \texttt{UST} needs to save necessary registers (context) beforehand, and then recover them before returning to the caller. 

\item \textbf{Record the duration and return ($38$ bytes):} \texttt{UST} will invoke \texttt{Tracer} again to record the timestamp after API invocation finishes. \texttt{UST} passes information required by the \texttt{Tracer} by pushing them onto the call stack. After \texttt{Tracer} finishes, \texttt{UST} will jump back to the real return address.

\end{itemize}

\subsection{Tracer}
\label{sec:tracer}

The \texttt{Tracer} component is responsible for tracing and collecting the information of each
API invocation by the \texttt{Universal Shadow Table}.

\Scaler{} keeps separate data for each thread. In this way, there is no need to utilize mutex locks for the update, as different threads are not updating the same tracing data simultaneously. Further, such a design reduces cache misses caused by true/false sharing~\cite{Sheriff}, but at the cost of more memory/storage consumption. The per-thread data will be stored in separate files in the end, and the \texttt{Offline Visualizer} will integrate all data from different threads together in the end. Note that \Scaler{} employs \texttt{initial-exec} TLS \- model~\cite{tlsmodel} to store per-thread variables, which only require a single \texttt{mov} instruction to access. In contrast, the default model (\texttt{dynamic} TLS) requires an extra function invocation called \texttt{\_\_tls\_get\_addr}, which is very inefficient. 
%\ST{which is not only inefficient but also complicated}. \ST{Remember to mention that Scaler can parse the tlsoffset and access the TLS variable in assembly code.}

To collect the execution time, \Scaler{} employs the light-weight \texttt{rdtsc} instruction that can read CPU clock cycles in user space efficiently, avoiding expensive system calls. It collects the timestamp before and after each API execution, and the difference between these two timestamps is the execution time of the current API invocation. The \texttt{Tracer} will invoke the \texttt{Online Data Folder} to record the collected information.

% \begin{enumerate} 
%     \item \Scaler{} saves necessary registers so that they can be used freely by the timing handler.
%     \item It collects the timestamp before invoking the API using \texttt{rtdsc} instruction. 
%     \item It invokes the real API directly, which is passed by the execution in Universal Shadow Table. 
%     \item It collects the timestamp again in order to compute the runtime of the API invocation. 
%     \item It recovers the saved registers, and jumps back to the original return address.
% \end{enumerate}

\subsection{Online Data Folder}
\label{sec:datafolder}

\Scaler{} designs a data structure that folds the traced data online. The data folding will follow two principles: (1) it reduces the memory/storage as much as possible, which also helps reduce the time of offline analysis. (2) It preserves the accuracy of traced data. In the end, The data folder outputs the traced data to the disk files at the end of the execution or upon receiving the signal from users. Each thread outputs one copy of data.
%With data folding, \Scaler{} 
%, and only outputs the traced data to the disk files in the end of the execution or receiving the signal from users

\paragraph{Efficient data folding without losing the accuracy:} 
\Scaler{} aims to report the time consumption and number of invocations for each API invoked by any component (library or application). We have the following \textbf{two observations} on API invocations: (1) the number of APIs that can be invoked by a component is constant, equaling to the total of its linkage table entries, and \texttt{dlsym} invocations, which remains unchanged at different execution time.
%Not good. \ST{How about changing to ``remains unchanged between different executions''?};  
%\todo{Steven: I think it is not correct to say that the dlsym invocation number will remain unchanged as time progresses.} 
(2) One API can be invoked by different components; For example, \texttt{pthread\_mutex\_lock} can be invoked by the application itself or different libraries. \Scaler{}'s design is built on these two observations. Based on the first observation, it utilizes an array-based structure %\todo{Steven: The data store uses an expandable array because new APIs may be invoked as time progresses and we need to expand the array to accommodate these new APIs. When the array expands, there is no need to perform synchronization because this array is thread-local.} 
to track the invocation information accumulatively for APIs, which could be determined by analyzing the corresponding elf files. That is, \Scaler{}'s memory/storage overhead is proportional to the number of linkage table entries and the number of \texttt{dlsym}-opened functions. Based on the second observation, \Scaler{} tracks the invocations of the same API by different components separately, preserving the accuracy. 

\paragraph{Handle abnormal program exits:} 
As mentioned above, \Scaler{} persists per-thread data to the disk when a thread exits. To intercept thread exits, the intuitive method is to replace the standard thread creation with a wrapper function so that it can invoke the real thread function inside and handle thread exits correspondingly. However, this method cannot intercept abnormal exits of threads. For instance, some programs may invoke pthread\_exit() explicitly to terminate threads (e.g., aget~\cite{aget}), and some children threads never exit (e.g., OpenMP). 
%, and \todo{OpenMP applications will never kill worker threads until the process finishes}. 
To handle these abnormal cases, \Scaler{} registers a common exiting handler via \texttt{\_\_cxa\_thread\_atexit}. For never-exiting threads, the main thread persists all remaining threads on their behalf when it exits. 

\paragraph{\FLOW{Attributing the runtime of invocations differently for serial and parallel phase:}} 
Obviously, an API invoked in the serial phase or in the parallel phase makes different performance impacts on the end-to-end performance. However, existing profilers, e.g., \texttt{perf}, typically summarize (and calculate the percentage) the execution time of different APIs, which cannot reveal the inherent performance issues of multithreaded programs inside~\cite{Coz}. \Scaler{} takes into account the difference for invocations occurring in serial and parallel phases. More specifically, in the recording phase, \Scaler{} divides the execution time of API invocation by the number of active threads in the parallel phase, and utilizes the original execution time for invocations in the serial phase. 

\subsection{Offline Visualizer}
\label{sec:visualizer}
\FLOW{\Scaler{}'s offline visualizer includes Python scripts used to generate component views and API views based on the log files generated by \Scaler{} runtime.} As discussed above, the offline visualizer integrates the per-thread data together. Since \Scaler{} already attributes the runtime of invocations for multithreaded programs as discussed in Section~\ref{sec:datafolder}, it simply summarizes the data from different threads together in its offline analysis.  

The component view shows the execution time for itself ( ``Self'') and other components. For instance, the component view for the application shows the percentage of runtime spent in the application and other libraries that are invoked directly by the application, while the component view for a library also includes the runtime spent in the library (``Self'') and other libraries called by this library. In particular, the runtime of ``Self'' equals the total runtime of this component minus the runtime spent on all APIs invoked by this component. For the runtime of a specific library, \Scaler{} simply summarizes the runtime of all APIs belonging to this library, which can be collected by analyzing the corresponding elf file. 
%\TP{The total time of the thread-creator component (usually the main application) equals the timestamp at the thread exit time minus the timestamp at thread creation time.}

%The component time reflects how much time the caller component invokes the callee component. To calculate per-thread component time, the offline visualizer sums API invocation time corresponding to the selected caller component and the same callee component. In the recorded log file, all components correspond to an incremental id, so the offline visualizer can just create an array and use the component id as an index to efficiently find data required for the summation.  To calculate per-thread ``Self'' time, the offline visualizer subtracts the component API invocation time from the total time of a component. The total time of the thread-creator component (usually the main application) equals the timestamp at the thread exit time minus the timestamp at thread creation time. The total time of the non-thread-creator component equals summation of API invocation time with the same callee component. The API invocation time of a component can be calculated by summing the API invocation time that corresponds to the same caller.

For the API view, \Scaler{} simply aggregates per-thread data together for all APIs. As mentioned in Section~\ref{sec:tracer}, each thread has a copy of the data that records API invocations by this thread. Since different threads employ the same array-based data structure for tracking invocations, \Scaler{} only needs to summarize the data with the same index together (indicating the same API) in different per-thread arrays.

%\FLOW{The offline visualizer do not need to perform extra calculation to display the ``API view'' and \TP{aggregate per-threaded data into one graph.}} The offline visualizer can read API invocation time directly from the recorded log file. This per-thread API invocation time time is already scaled by the multi-threaded runtime attribution at runtime, so the offline visualizer simply need to add up the scaled time for each thread together when displaying data.

Note that \Scaler{}'s component view displays the waiting time (shown as ``Wait'') separately instead of counting them as 
normal API invocations of \texttt{pthreads} library.  The API invocations related to the waiting (e.g., condition/barrier waits) indicate that programs are not actively doing useful work, leading to serious performance issues. Separating the waiting time into a distinct category helps identify such issues as mentioned in Section~\ref{sec:effectiveness}. Further, \Scaler{} summarizes the waiting time from different types of threads together, reporting load imbalance issues when different types of threads have significantly different amounts of waiting time. \SC{} learns such a mechanism from SyncPerf~\cite{SyncPerf}. Note that \SC{} is able to achieve this due to its capability of the full trace.

%\FLOW{\Scaler{} displays thread join and wait times separately.} Since long execution time in wait and join typically means the caller of \texttt{pthread} library misused these APIs rather than the poor implementation of these APIs by \texttt{pthread}. 

%% file: src/4_evaluation.tex
\section{Evaluation \label{sec:Evaluation}}

% \ST{Reviewer think we should use more factual statements like rank of info, how percentage usage compares to other usages in the visualizations .etc. Should keep more factual statements like rank of info, how percentage info.}

% \ST{We should state clearly the purpose of experiments. Self-constructed example proves implementation correctness. Effectiveness example proves usefulness. Reviewers have question}

% % \ST{Compare with popular trace-based tool. PIN, DynamoRIO}
% \ST{Steven think that we need to add a section comparing perf's output and application's output}.

This section aims to answer the following research questions, by comparing with other existing work:

\begin{itemize}
    \item Can \Scaler{} detect some performance bugs? (Section~\ref{sec:effectiveness})
    \item What is the runtime overhead of \Scaler{}? (Section~\ref{sec:performanceOverhead})
    \item What is the memory overhead of \Scaler{}? (Section~\ref{sec:memoryOverhead})
    % \item Does the proposed synchronization method in Multithreaded Runtime Attribution work correctly?  (Section \ref{sec:EffectOfMultiThreadedRuntimeAttribution})
\end{itemize}

\subsection{Experimental Setting \label{sec:ExperimentalSetting}}

\paragraph{\FLOW{Hardware/Software Platform}} \FLOW{We performed all experiments on a machine with dual $20$-core,$40$-hyperthread Intel® Xeon® Gold 6230 CPUs, installed with 256GB memory.} The system version is Ubuntu 18.04.6, with the kernel version 5.4.0-146-generic. The compiler used is \texttt{gcc/g++} 7.5.0-3, with the optimization level ``-O3''. 

\paragraph{Evaluated Applications:} \FLOW{We chose \parsec{} benchmark suite~\cite{bienia2008parsec} version 3.0-20150206, including $13$ multithreaded programs, and multiple real-world applications, including \texttt{mem\-cach\-ed}-1.6.17, \texttt{MySQL-8.0.31}, \texttt{nginx-1.23.2}, \texttt{Redis\--7\-.0\-.4}, and \texttt{SQLite}-3.39.4.} For the evaluation, most \parsec{} applications (except \texttt{dedup} and \texttt{ferret}) use $80$ threads and the default parameters. Both \texttt{dedup} and \texttt{ferret} use 16 threads for each pipeline stage, with 51 and 67 threads in total.  For \texttt{Memcached}, we use the \texttt{memtier-1.4.0} client that runs for $60$ seconds. For \texttt{MySQL}, we use \texttt{sysbench-1.0.20} client to run oltp\_read\_write test for 10 tables of size $10000$. For \texttt{nginx}, we use the \texttt{wrk-4.1.0.3} client that runs $10,240$ connections for 60 seconds. For \texttt{Redis}, we use \texttt{redis\--benchmark\--5.6.0.16} to send $100,000$ requests in total. For \texttt{SQLite}, we use \texttt{threadtest3} for the evaluation. 

\subsection{Effectiveness Evaluation of \Scaler{} \label{sec:effectiveness}}

\begin{table*}[ht!]
\setlength{\tabcolsep}{2pt}
\scalebox{0.75}{
\begin{tabular}{lllcccc}
    \hline
    BugId & Abnormal Behavior & Root Cause & \perf{} & \Scaler{} & New Bug & Speedup \\ \hline
    \texttt{canneal} & Extensive string comparisons & Improper data structure of application & \xmark & \cmark & \cmark & $51.6\%$ \\
    \texttt{dedup}-1 & Extensive time on read()/write() & Inefficient I/O operations of application & \xmark & \cmark & \cmark & $49.1\%$ \\
    \texttt{dedup}-2 & Imbalance waiting time & Improper thread assignment of application & \xmark & \cmark & \xmark & $25.5\%$ \\
    \texttt{dedup}-3 & Extensive \texttt{madvise} calls & Improper configuration of library & \xmark & \cmark & \xmark & $74.2\%$ \\
    \texttt{ferret} & Imbalance waiting time & Improper thread assignment of application & \xmark & \cmark & \xmark & $76.7\%$ \\
    \texttt{swaptions} & Significant lock time & Improper configuration of library & \cmark & \cmark & \xmark & $44.0\%$ \\ \hline
    \end{tabular}
}
\caption{Effectiveness comparison between \Scaler{} and \perf{}. \label{tbl:EffectivenessSummary} }
\end{table*}

% \ST{Move table 1 to introduction.}
% \ST{Compare with add compiler based tools here and say it is infeasible to work in this case.}

We evaluated all the above-mentioned applications to confirm \Scaler{}'s effectiveness and compared the result with \perf{}. Such comparison shows the necessity of our proposed lightweight full-trace method. We only show the comparison with \perf{} due to the following reasons: first, \perf{} does not require recompilation of the application and libraries, which is as convenient as \SC{}. Second, other full trace based profilers have significantly higher runtime overhead compared to \Scaler{}, making it unnecessary for the comparison.  Instead, \perf{}, a sampling based profiler,  has a similar performance overhead as \Scaler{}. Other sampling based tools have both lower collection frequency and higher runtime overhead compared to \perf{}. 
%Additionally, comparing effectiveness with a sampling-based profiler highlights the necessity of our proposed lightweight full-trace method. 

Overall, \Scaler{} detects $6$ bugs in highly-optimized benchmarks, as shown in Table~\ref{tbl:EffectivenessSummary},  while \perf{} can only detect one bug. Note that $2$ out of these $6$ bugs are reported for the first time. After fixing these bugs, the performance improves between $25.5\%$ and $76.7\%$ (as shown in the Column ``Speedup''). Among these bugs, $4$ bugs are related to the design and implementation of the main application, indicating that \textit{the abnormal behavior of APIs can be utilized to infer the inappropriate design or configuration of the application}. The other two performance bugs (\texttt{dedup}-3 and \texttt{swaptions}) are related to performance issues caused by external libraries. The results show the effectiveness of \Scaler{}.
Note that \Scaler{} may miss bugs caused by non-API functions or hardware, as discussed in Section~\ref{sec:limitation}. 

In the remainder of this section, we will study performance issues listed in Table~\ref{tbl:EffectivenessSummary}. Since we already discussed the \texttt{canneal} bug in Section~\ref{sec:Introduction} and Figure~\ref{fig:intro_xfa_basicidea_b}, this example will be skipped.
Further, since the root cause of \texttt{dedup-2}~\cite{SyncPerf} and \texttt{ferret} is similar, we only describe the \texttt{ferret} example here.

\subsubsection{Case Study 1: Inefficient I/O Operations of \texttt{dedup} \label{sec:eval_effectiveness_dedup_malloc}}

\SC{} reports a \textbf{new} performance bug (\texttt{dedup-1}) in \texttt{dedup}, which de-duplicates and compresses files through a four-stage pipeline, including fragmentation, deduplication, compression, and data reordering~\cite{bienia2008parsec}.  Figure~\ref{fig:eval_effectiveness_scaler_dedup3_output} shows the report of \SC{}: the application spends $35\%$ time on the \texttt{glibc} library, which is even higher than $33\%$ of ``Self''; In the API view of the \texttt{glibc} library,  \texttt{read} accounts for $18\%$ of the total execution time and \texttt{write} accounts for $35\%$ of the execution time; Further, \Scaler{} reports that \texttt{write} was invoked $1,109,852$ times in total. From such a report, we could infer that the application invokes extensive \texttt{read}() and \texttt{write}() to process large files. In fact, a more efficient alternative is to utilize the \texttt{mmap} API to map files to the memory and then operate on it directly with pointers, since the alternative will eliminate the overhead of system calls, and  leverage page caching and COW semantics of the memory system. After changing to the method of using \texttt{mmap}, the performance is improve by $49.1\%$. In contrast, \perf{} cannot find this bug,  because it only reports that $4.12\%$ of time is spent on \texttt{write()}, possibly due to its coarse sampling rate.

% =================================================
% The following format are used in dual column papers
% =================================================
% \begin{figure}
%      \centering
%      \includegraphics[width=2.4in]{src/figures/eval_effectiveness_scaler_dedup1_output.pdf}
%      \caption{\Scaler{}'s output for \texttt{dedup-1}. \label{fig:eval_effectiveness_scaler_dedup1_output}}
% \end{figure}

% \begin{figure}
%      \centering
%      \includegraphics[width=2.4in]{src/figures/eval_effectiveness_perf_dedup1_output.pdf}
%      \caption{\perf{}'s output for \texttt{dedup-1}. \label{fig:eval_effectiveness_perf_dedup1_output}}
% \end{figure}

% =================================================
% The following format are used in single column papers
% =================================================

\begin{figure}[ht!]
     \centering
     \begin{subfigure}[b]{0.45\textwidth}
        \centering
        \includegraphics[width=2.5in]{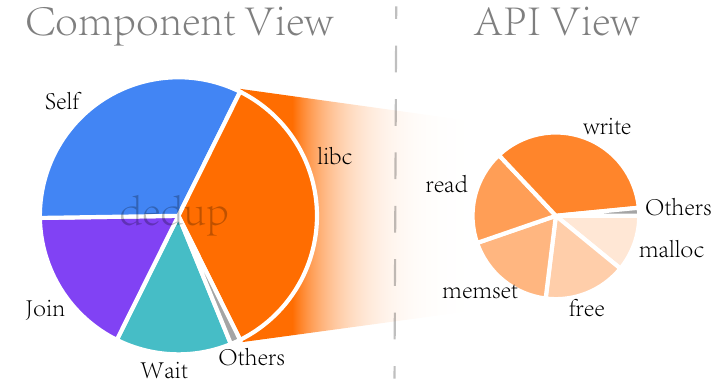}
        \caption{\Scaler{}'s output.\label{fig:eval_effectiveness_scaler_dedup1_output}}
    \end{subfigure}
    \hfill
    \begin{subfigure}[b]{0.45\textwidth}
        \centering
        \includegraphics[width=2.5in]{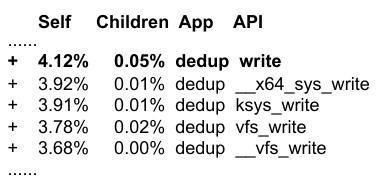}
        \caption{\perf{}'s output.} 
    \end{subfigure}
    \Description[Profiling output for \texttt{dedup-1}.]{Profiling output for \texttt{dedup-1}.}
    \caption{Profiling output for \texttt{dedup-1}.}
\end{figure}

\subsubsection{Case Study 2: Extensive \texttt{madvise} Invocations of \texttt{dedup}} 

\begin{figure}[ht!]
    \centering
    \begin{subfigure}[b]{0.45\textwidth}
        \centering
        \includegraphics[width=2.5in]{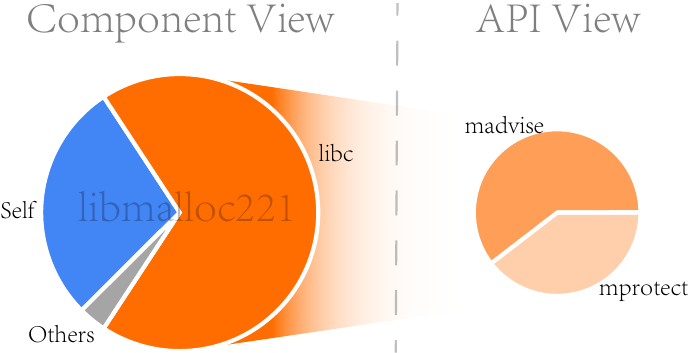}
        \caption{\Scaler{}'s output.}
        \label{fig:eval_effectiveness_scaler_dedup3_output}
    \end{subfigure}
    \hfill
    \begin{subfigure}[b]{0.45\textwidth}
        \centering
        \includegraphics[width=2.5in]{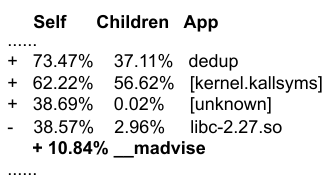}
        \caption{\perf{}'s output.} 
    \end{subfigure}
    \Description[Profiling output for \texttt{dedup-3}.]{Profiling output for \texttt{dedup-3}.}
    
    \caption{Profiling output for \texttt{dedup-3}.}
\end{figure}

%\todo{What do you mean libmalloc221? We should change it to the allocator of \texttt{glibc-2.21}. Also, PLEASE remove "libmalloc221" from the Figure, it is ugly and misleading. }
\texttt{dedup} has another known performance problem when linked with the memory allocator of \texttt{glibc-2.21}~\cite{madvise}. For this case, \Scaler{}'s output can be seen in Figure \ref{fig:eval_effectiveness_scaler_dedup3_output}. The allocator spends 68\% time on two APIs, e.g.,  \texttt{madvise} and \texttt{mprotect}. The underlying reason for this issue is that the allocator frequently releases the allocated virtual memory (\texttt{madvise}) back to the OS based on a threshold. But \texttt{madvise} needs to acquire the per-process lock of protecting memory regions inside the OS, and introduces extensive page faults when such a region is accessed again. The acquisitions of the lock introduce high kernel contention with page faults and other memory-related system calls (e.g., \texttt{mprotect}).  
%\texttt{madvise} releases allocated virtual memory back to the OS. However, frequent \texttt{madvise} calls of \texttt{libmalloc221} introduces high kernel contention in page faults and memory-related system calls. 
By increasing the threshold (via the configuration), we can significantly reduce the amount of \texttt{madvise} invocations and kernel contention correspondingly. The fix improved the performance by 74.2\%. \perf{} cannot reveal this bug, since \texttt{madvise} only accounts for $10.84\%$ time, which is even lower than $17.66\%$ of \texttt{memset}. 

\subsubsection{Case Study 3: Thread Imbalance of \texttt{ferret}}

\begin{figure}[ht!]
     \centering
     \begin{subfigure}[b]{0.45\linewidth}
         \centering
         \includegraphics[width=1in]{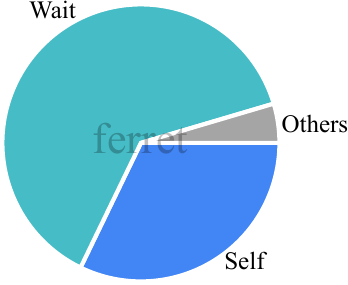}
         \caption{Wait time exceeds execution time}
         \label{fig:eval_effectiveness_ferret_scaler_imbalance_output_a}
     \end{subfigure}
     % \hfill
     \begin{subfigure}[b]{0.45\linewidth}
         \centering
         \includegraphics[width=1.5in]{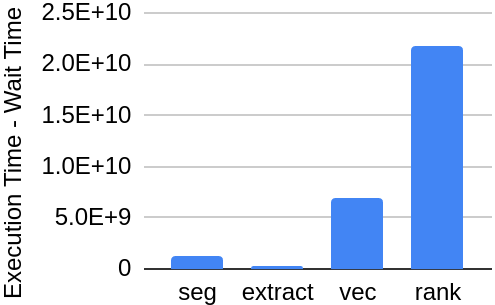}
         \caption{Imbalance of thread execution time}
         \label{fig:eval_effectiveness_ferret_scaler_imbalance_output_b}
     \end{subfigure}
     \caption{\Scaler{}'s output for \texttt{ferret}}
     \Description[\Scaler{}'s output for \texttt{ferret}]{\Scaler{}'s output for \texttt{ferret}}
\end{figure}

\texttt{Scal\-er} reports a thread-imbalance problem in \texttt{ferret} that implements a content-based similarity search:  Figure~\ref{fig:eval_effectiveness_ferret_scaler_imbalance_output_a} shows that \texttt{ferret} spends over 50\% time on the waiting, and Figure~\ref{fig:eval_effectiveness_ferret_scaler_imbalance_output_b} further reveals that different thread groups have different effective execution time, where \texttt{rank}'s effective execution time is about $16\times$ higher than that of \texttt{seg}. Therefore, this is a clear indication of thread-imbalance problem, as observed by SyncPerf~\cite{SyncPerf}. We fixed this issue by adjusting the thread assignment from the default $16$:$16$:$16$:$16$ (related to \texttt{seg}:\texttt{extract}:\texttt{vec}:\texttt{rank} threads) to $3$:$1$:$15$:$45$. This fix improves the performance by $76.7\%$. In contrast, \perf{}'s report cannot reveal this problem, since it does not summarize the runtime of different thread groups together and its too-coarse sampling mechanism.  
%due to the inaccuracy and the lack of proper thread imbalance data attribution.

\subsubsection{Case Study 4: Improper Configuration of The \texttt{hoard} Allocator}

\texttt{swaptions} of \parsec{} has a known performance bug, caused by the improper configuration of the \texttt{hoard} allocator~\cite{CachePerf}.  The component view of \texttt{libhoard} shows that it spends $93\%$ time on the \texttt{pthread} library. The API view of the \texttt{pthread} shows that the spin lock takes $99\%$ of the time. Therefore, this is a clear lock contention issue caused by the allocator. We can fix this issue by changing the \texttt{SUPERBLOCK\_SIZE} flag from $4096$ to $65536$ and \texttt{EMPT\-INESS\_CLASSES} from 8 to 2. After the fix, the performance improved by $44\%$. Note that \perf{} can also reveal this bug, since it also reports the large percentage of time is spent inside the spin lock of \texttt{libhoard}.

% \FLOW{This example has been analyzed in Section \ref{sec:Introduction}.} \Scaler{}'s output can be seen in Figure \ref{fig:intro_xfa_basicidea_b}. It is abnormal for an allocator to spend lots of time on locks. Since this performance slowdown is so severe, \perf{} can also find this bug even though it has a low sampling rate. This is an existing performance bug, and we reproduced and confirmed that it originates from the improper configuration of the \texttt{Hoard} allocator. 

\subsection{Performance Overhead of \Scaler{} \label{sec:performanceOverhead}}

\begin{table}
% \centering
\scalebox{0.65}{
\begin{tabular}{clrrrrr}
    \hline
    \multicolumn{1}{l}{Category} & Application & \multicolumn{1}{c}{\perf{}} & \multicolumn{1}{c}{\vtuneums{}} & \multicolumn{1}{c}{\bpftrace{}} & \multicolumn{1}{c}{\ltrace{}} & \multicolumn{1}{c}{\Scaler{}} \\ \hline
    \multirow{13}{*}{\parsec{}} & \texttt{blackscholes} & $16.3\%$ & $32.2\%$ & $183.0\times$ & $>4953.8\times$ & $35.0\%$ \\
     & \texttt{bodytrack} & $21.3\%$ & $28.4\%$ & $34.0\times$ & N/A & $16.1\%$ \\
     & \texttt{canneal} & $12.9\%$ & $12.6\%$ & $14.8\times$ & N/A & $79.3\%$ \\
     & \texttt{dedup} & $31.8\%$ & $73.0\%$ & $1.4\times$ & N/A & $6.2\%$ \\
     & \texttt{facesim} & $11.8\%$ & $20.8\%$ & $26.3\times$ & N/A & $15.8\%$ \\
     & \texttt{ferret} & $9.8\%$ & $17.5\%$ & $43.6\times$ & N/A & $7.9\%$ \\
     & \texttt{fluidanimate} & $16.0\%$ & $24.5\%$ & $174.8\times$ & N/A & $37.8\%$ \\
     & \texttt{freqmine} & $18.2\%$ & $35.4\%$ & $15.0\times$ & N/A & $28.6\%$ \\
     & \texttt{raytrace} & $3.2\%$ & $5.2\%$ & $5.5\times$ & N/A & $10.1\%$ \\
     & \texttt{streamcluster} & $44.7\%$ & $37.9\%$ & $5.3\times$ & N/A & $14.4\%$ \\
     & \texttt{swaptions} & $31.6\%$ & $56.7\%$ & $4.4\times$ & $>7437.7\times$ & $33.8\%$ \\
     & \texttt{vips} & $1.3\times$ & $1.8\times$ & $2.4\times$ & N/A & -$0.6\%$ \\
     & \texttt{x264} & $37.5\%$ & $1.7\times$ & $4.1\%$ & N/A & $14.3\%$ \\ \hline
    \multirow{5}{*}{\begin{tabular}[c]{@{}c@{}}Real-world\\ Application\end{tabular}} & \texttt{memcached-1.6.17} & $0.8\%$ & $2.1\%$ & $5.5\times$ & N/A & $20.0\%$ \\
     & \texttt{MySQL-8.0.31} & $4.5\%$ & $1.7\%$ & $1.8\times$ & N/A & $0.1\%$ \\
     & \texttt{nginx-1.23.2} & $5.3\%$ & $30.9\%$ & $59.7\%$ & N/A & $5.1\%$ \\
     & \texttt{Redis-7.0.4} & $9.4\%$ & $9.8\%$ & $46.7\%$ & N/A & $20.5\%$ \\
     & \texttt{SQLite-3.39.4} & $2.0\%$ & $0.8\%$ & $48.6\%$ & N/A & $21.9\%$ \\ \hline
    \multicolumn{1}{l}{\textbf{Overhead}} & - & $22.5\%$ & $41.4\%$ & $28.8\times$ & $>6195.7\times$ & $20.3\%$ \\ \hline
\end{tabular}
}
\caption{Performance overhead of \Scaler{} and others. \label{tbl:performance}}
\end{table}
% \ST{Do not just report the numbers again, but the reader expect an interpretation of results and conclusions.}
% \subsubsection{Comparison with Existing Work} 
In this section, we further evaluate the performance overhead of \Scaler{}, and compare it with existing work. For \perf{}, we use sampling rate $4000$ and the ``-g'' option (in order to collect the callstack). For \vtune{}, we use the default sampling interval -- 10ms and select the default ``User Mode Sampling (ums)'' option. For \bpftrace{}, we use \Scaler{} to collect the list of all invoked APIs, and then write a \texttt{bpftrace} script to attach \texttt{uprobe} and \texttt{uretprobes} for all of these APIs. For \ltrace{}, we use ``--no-signals -o /dev/null'' flag to minimize the impact of signal and output. 

%\todo{What is the difference of CNT, TIME, ATTR? Why ATTR is much more than Timing? Whether TIME incudes both CNT and TIME? } \ST{CNT refers to the counting mode, TIME refers to the timing mode, and ATTR refers to the multithreaded runtime attribution mode. In \Scaler{}'s implementation, multithreaded runtime attribution mode only increases timing mode overhead by 3.8\%. 

\subsubsection{Overhead of Online Tracing}

Table~\ref{tbl:performance} shows the performance overhead of \Scaler{} and other existing work. We evaluate \Scaler{}, \perf{} and \vtuneums{} all run $8$ times and we report the average. \bpftrace{} introduce daunting performance overhead, so we can only test it once. We randomly picked two applications in \parsec{} to profile with \ltrace{} and both cannot finish within $24$ hours. We also observed freezes on real-world applications when profiling with \ltrace{}. Consequently, we did not test \ltrace{} for all programs, since two evaluated programs all require more than $24$ hours to finish. Overall, \Scaler{} introduces $20.3\%$ performance overhead, which is the least among all  evaluated tools. 

\Scaler{}'s lower overhead can be attributed to its efficient internal design, which includes the selective binary instrumentation method that minimizes binary instrumentation, the Universal Shadow Table that minimizes data attribution overhead, and the pure user-space profiling without involving the context switch overhead. In comparison, both \bpftrace{} and \ltrace{} employ the software breakpoint technique (e.g., INT3) that requires saving and restoring the user context for each API invocation. The difference is that \ltrace{} involves context switching between the \ltrace{} tool,  the traced program, and the kernel while \bpftrace{} uses \ebpf{} to process data directly in the kernel to avoid the context switch between kernel space and user space. Even with in-kernel processing, the context switch overhead still imposes high overhead due to the extensive number of API invocations (around $62.9$ million each second as shown in Table~\ref{tbl:SamplingRateComparison}).

For sampling-based profilers, \perf{}'s overhead is around $22.5\%$, and \vtuneums{}'s overhead is around $41.4\%$. Both of them impose a higher overhead than \SC{}, even \SC{} provides a full trace functionality and collects orders of magnitude more events. \perf{} utilizes the hardware Performance Monitoring Units (PMUs) to sample the execution status, and collects the callstack at each sample. Similarly, it also involves context switches between user space and kernel space. \vtuneums{} is also a sampling-based tool but with the default sampling rate that is $10\times$ lower than \perf{}. The underlying reason for its high overhead is that it utilizes the slow \texttt{ptrace} system call to collect the information.

\paragraph{Difference of Recorded Events}

\begin{table}
\centering
\footnotesize
\scalebox{0.85}{
\begin{tabular}{clrr}
\hline
\multicolumn{1}{l}{Category} & Application & \multicolumn{1}{c}{\begin{tabular}[c]{@{}c@{}}\Scaler{} \\ (Baseline)\end{tabular}} & \multicolumn{1}{c}{\perf{}} \\ \hline
\multirow{13}{*}{\parsec{}} & \texttt{blackscholes} & 5.02E+09 & 1.13E+06 \\
 & \texttt{bodytrack} & 2.23E+09 & 5.28E+06 \\
 & \texttt{canneal} & 1.20E+09 & 1.75E+06 \\
 & \texttt{dedup} & 1.19E+07 & 2.63E+05 \\
 & \texttt{facesim} & 4.40E+09 & 1.55E+07 \\
 & \texttt{ferret} & 2.17E+09 & 3.40E+06 \\
 & \texttt{fluidanimate} & 1.35E+10 & 1.09E+07 \\
 & \texttt{freqmine} & 2.60E+08 & 2.27E+06 \\
 & \texttt{raytrace} & 1.67E+09 & 3.17E+06 \\
 & \texttt{streamcluster} & 3.11E+08 & 2.79E+07 \\
 & \texttt{swaptions} & 3.74E+09 & 2.90E+06 \\
 & \texttt{vips} & 2.36E+07 & 1.72E+06 \\
 & \texttt{x264} & 3.10E+05 & 4.28E+05 \\ \hline
\multirow{5}{*}{\begin{tabular}[c]{@{}c@{}}Real-world \\ Applications\end{tabular}} & \texttt{memcached-1.6.17} & 3.83E+08 & 9.34E+05 \\
 & \texttt{MySQL-8.0.31} & 6.85E+08 & 1.84E+06 \\
 & \texttt{nginx-1.23.2} & 8.02E+03 & 2.38E+05 \\
 & \texttt{Redis-7.0.4} & 4.77E+08 & 9.23E+04 \\
 & \texttt{SQLite-3.39.4} & 3.51E+08 & 1.46E+05 \\ \hline
\multicolumn{1}{l}{\textbf{Avg Counts}} & - & 2.03E+09 & 4.44E+06 \\ \hline
\multicolumn{1}{l}{\textbf{Avg Freq}} & - & 6.29E+07 & 1.05E+05 \\ \hline
\end{tabular}
}
\captionsetup{justification=justified}
\caption{Number of events recorded by  \Scaler{} and \perf{}.\label{tbl:SamplingRateComparison}} 
\end{table}

\Scaler{} provides the full trace functionality, collecting a significantly higher number of events than sampling-based tools, e.g., \perf{}. Table~\ref{tbl:SamplingRateComparison} shows the difference between \Scaler{} and \perf{}. On average, \perf{} only records 105 thousand events each second, while \Scaler{} records $62.9$ million events each second. That is, \Scaler{} collects data $599\times$ more frequently than \perf{}, while imposing less performance overhead.

\subsubsection{Overhead of Offline Analysis}

We further compared the performance overhead of offline analysis between \Scaler{} and \perf{}. \Scaler{}'s offline visualizer (writing with Python) only takes an average of $0.43$ seconds to execute, while \perf{}'s offline analysis takes 33.3 seconds on average. That is, \perf{}'s offline analysis is around $76\times$ slower than \Scaler{}. The underlying reason is that \Scaler{} performs the majority of computation online via its relation-aware data folding. 
%and the visualization tool only reads data and provides user-friendly print output, so \Scaler{} only incurs an average of 0.43 seconds post-processing time even if we implemented the visualization tool in Python. The fast visualization also results from the efficient data structure mentioned in Section \ref{sec:DesignAndImpl} because of the small data recording size and input scale irrelevance.
In contrast, \perf{} saves the call stack of every sample and aggregates the recorded events offline.   %\perf{}'s C/C++ implemented visualization tool incurs an average of 33.35 seconds post-processing time, around $76\times$ slower compared to \Scaler{}.

\subsection{Memory Overhead of \Scaler{} \label{sec:memoryOverhead}}

We also evaluated the memory overhead of \Scaler{}, \perf{}~\cite{perf}, \vtuneums{}~\cite{vtune-ums} and \bpftrace{}~\cite{bpftrace}. The results can be seen in Table \ref{tbl:MemoryOverhead}. Overall, \Scaler{} introduces orders of magnitude lower memory overhead compared to other existing work.  Specifically, \Scaler{} only imposes $15.5\%$ memory overhead. In comparison, \perf{}'s memory overhead is $6.7\times$, \vtuneums{}'s overhead is around $18.3\times$, and \bpftrace{}'s overhead is  $3.1\times$. We observe that \perf{} and \vtuneums{} has relatively high memory overhead for application with small memory footprints, while \Scaler{} still provides low memory overhead for these applications. The major reason for \Scaler{}'s low overhead is due to its Relations-Aware Data Folding. The major memory overhead for \Scaler{} is proportional to the number of APIs. For each API, \Scaler{} allocates one universal shadow table ($134$ bytes) entry and one struct that records API-specific information ($112$ bytes). All runtime data are dynamically folded at runtime and do not need extra space for recording.

\begin{table}
\centering
\footnotesize
\scalebox{0.95}{
\setlength{\tabcolsep}{1pt}
\begin{tabular}{clrrrrr}
    \hline
    \multicolumn{1}{l}{Category} & Application & \multicolumn{1}{c}{Original} & \multicolumn{1}{c}{\perf{}} & \multicolumn{1}{c}{\vtuneums{}} & \multicolumn{1}{c}{\bpftrace{}} & \multicolumn{1}{c}{\Scaler{}} \\ \hline
    \multirow{13}{*}{\parsec{}} & \texttt{blackscholes} & 617MB & $1.3\%$ & $8.0\%$ & $17.2\%$ & $2.6\%$ \\
     & \texttt{bodytrack} & 43MB & $16.1\times$ & $16.1\times$ & $2.5\times$ & $11.9\%$ \\
     & \texttt{canneal} & 858MB & $1.0\%$ & $8.4\%$ & $12.4\%$ & $1.6\%$ \\
     & \texttt{dedup} & 1564MB & -$0.7\%$ & $10.3\%$ & $13.3\%$ & $8.0\%$ \\
     & \texttt{facesim} & 337MB & $4.0\times$ & $1.4\times$ & $49.5\%$ & $3.2\%$ \\
     & \texttt{ferret} & 142MB & $2.0\times$ & $4.0\times$ & $75.5\%$ & $6.2\%$ \\
     & \texttt{fluidanimate} & 1020MB & $4.1\%$ & $11.0\%$ & $10.4\%$ & $1.1\%$ \\
     & \texttt{freqmine} & 3388MB & $0.8\%$ & $2.5\%$ & $4.5\%$ & $0.8\%$ \\
     & \texttt{raytrace} & 1292MB & $0.7\%$ & $5.5\%$ & $8.2\%$ & $0.2\%$ \\
     & \texttt{streamcluster} & 116MB & $20.9\times$ & $5.6\times$ & $92.2\%$ & $25.2\%$ \\
     & \texttt{swaptions} & 13MB & $24.9\times$ & $49.0\times$ & $8.3\times$ & $81.8\%$ \\
     & \texttt{vips} & 225MB & $3.6\times$ & $2.0\times$ & $41.5\%$ & $4.9\%$ \\
     & \texttt{x264} & 1792MB & $0.3\%$ & $3.5\%$ & $6.0\%$ & $2.6\%$ \\ \hline
    \multirow{5}{*}{\begin{tabular}[c]{@{}c@{}}Real-world\\ Application\end{tabular}} & \texttt{memcached-1.6.17} & 7MB & $34.3\times$ & $88.3\times$ & $15.4\times$ & $62.9\%$ \\
     & \texttt{MySQL-8.0.31} & 666MB & $1.2\%$ & $2.8\times$ & $17.2\%$ & $3.7\%$ \\
     & \texttt{nginx-1.23.2} & 8MB & $9.1\times$ & $78.8\times$ & $12.9\times$ & $25.0\%$ \\
     & \texttt{Redis-7.0.4} & 12MB & $3.4\times$ & $50.9\times$ & $8.6\times$ & $16.9\%$ \\
     & \texttt{SQLite-3.39.4} & 21MB & $1.6\times$ & $29.3\times$ & $4.9\times$ & $21.0\%$ \\ \hline
    \multicolumn{1}{l}{\textbf{Overhead}} & - & - & $6.7\times$ & $18.3\times$ & $3.1\times$ & $15.5\%$ \\ \hline
\end{tabular}
}
\caption{Memory overhead of \Scaler{} and others.}
\label{tbl:MemoryOverhead}
\end{table}

\subsection{\revision{Sampling rate of \perf{}} \label{sec:Perf samplingrate}}
\revision{We further evaluate the impact of \perf{} sampling rate on the profiling result and the runtime overhead. In theory, setting a higher sampling rate will improve profiling accuracy at the cost of higher runtime overhead. However, in practice \perf{}'s sampling rate has limitations and will not strictly follow the frequency specified by the user. In Table \ref{tbl:PerfSamplingRateComparison}, we increased the sampling rate from 4000Hz used in Section \ref{sec:Evaluation} to 8000Hz and profiled programs under two different sampling rates. We also adjusted the kernel flag ``\texttt{perf\-\_event\-\_max\-\_\-sample\-\_rate}'' accordingly to increase the system sampling rate limit. We then used the official ``perf diff'' tool to calculate the time difference for every function recorded under different sampling rates. This tool normalizes the difference value to $0\%$-$100\%$ and output a percentage value in the report\cite{perfdiffdoc}. Finally, we parsed the report and find the function with the maximum difference value and reportd in column ``Max output difference''. From Table \ref{tbl:PerfSamplingRateComparison}, we can see that using a $2\times$ higher sampling rate on average only changed the maximum output difference by $0.57\%$. And the average performance overhead remains the same. This result indicates that using sampling rate 4000Hz in Section \ref{sec:Evaluation} is appropriate because it already reached the maximum feasible sampling rate in our experiment environment. 
}

\begin{table}
\centering
\footnotesize
\scalebox{0.85}{
\begin{tabular}{clrrr}
\hline
\multicolumn{1}{l}{Category} & Application & \multicolumn{1}{c}{\begin{tabular}[c]{@{}c@{}}perf-8000Hz\\ overhead\end{tabular}} & \multicolumn{1}{c}{\begin{tabular}[c]{@{}c@{}}perf-4000Hz\\ overhead\end{tabular}} & \multicolumn{1}{c}{\begin{tabular}[c]{@{}c@{}}Max output \\ difference\end{tabular}} \\ \hline
\multirow{13}{*}{\parsec{}} & \texttt{blackscholes} & $18.1\%$ & $16.3\%$ & $0.21\%$ \\
 & \texttt{bodytrack} & $45.7\%$ & $21.3\%$ & $0.15\%$ \\
 & \texttt{canneal} & $15.0\%$ & $12.9\%$ & $0.18\%$ \\
 & \texttt{dedup} & $7.9\%$ & $31.8\%$ & $0.40\%$ \\
 & \texttt{facesim} & $19.1\%$ & $11.8\%$ & $0.05\%$ \\
 & \texttt{ferret} & $11.2\%$ & $9.8\%$ & $0.27\%$ \\
 & \texttt{fluidanimate} & $17.0\%$ & $16.0\%$ & $0.09\%$ \\
 & \texttt{freqmine} & $21.3\%$ & $18.2\%$ & $0.32\%$ \\
 & \texttt{raytrace} & $4.5\%$ & $3.2\%$ & $1.33\%$ \\
 & \texttt{streamcluster} & $69.5\%$ & $44.7\%$ & $2.24\%$ \\
 & \texttt{swaptions} & $22.9\%$ & $31.6\%$ & $0.08\%$ \\
 & \texttt{vips} & $83.2\%$ & $1.3\times$ & $0.41\%$ \\
 & \texttt{x264} & $33.0\%$ & $37.5\%$ & $2.57\%$ \\ \hline
\multirow{5}{*}{\begin{tabular}[c]{@{}c@{}}Real-world \\ Applications\end{tabular}} & \texttt{memcached-1.6.17} & $3.4\%$ & $0.8\%$ & $0.51\%$ \\
 & \texttt{mysql-8.0.31} & $9.9\%$ & $4.5\%$ & $0.39\%$ \\
 & \texttt{nginx-1.23.2} & $14.6\%$ & $5.3\%$ & $0.15\%$ \\
 & \texttt{redis-7.0.4} & $1.8\%$ & $9.4\%$ & $0.71\%$ \\
 & \texttt{sqlite-3.39.4} & $2.9\%$ & $2.0\%$ & $0.14\%$ \\ \hline
\multicolumn{1}{l}{\textbf{Avg Overhead}} & - & $22.3\%$ & $22.5\%$ & - \\ \hline
\multicolumn{1}{l}{\textbf{Avg Output Diff}} & - & - & - & $0.57\%$ \\ \hline
\end{tabular}
}
\captionsetup{justification=justified}
\caption{\revision{The impact of increasing the sampling rate on the output and performance for \perf{}}.\label{tbl:PerfSamplingRateComparison}} 
\end{table}

\subsection{\revision{Corner case analysis}}

\revision{In the remainder of this section, we will analyze several corner cases to help facilitate the understanding of how \Scaler{} components work together to overcome these challenges.}

\subsubsection{\revision{Case Study 1: Some APIs are invoked by the pthread library before the recording context is initialized.}}

\revision{
\texttt{pthread} library will call \texttt{malloc} to allocate the memory required to construct the newly created thread. These \texttt{malloc} API invocations will still be intercepted because \Scaler{} has replaced the \texttt{.plt} of the pthread library at an earlier time. However, at this time \Scaler{}'s per-thread recording context will remain unallocated until the thread construction completes. \Scaler{} can distinguish these pre-mature API invocations using the first 20 bytes of assembly code in Universal Shadow Table as mentioned in Section \ref{sec:shadowtable}. After the recording context has been initialized, all subsequent \texttt{malloc}s invoked by the pthread library can still be recorded and thus skipping a few events during thread construction will not significantly change the profiling result.
}
\subsubsection{\revision{Case Study 2: \texttt{libstdc++6.0.32} uses ``jmp'' instruction to invoke APIs.}}

\revision{
Library such as \texttt{libstdc++6.0.32} uses a single ``jmp'' instruction in the \texttt{.plt} entry for ``operator delete(void*)'', which is different from the regular \texttt{.plt} introduced in Section \ref{sec:background}. \Scaler{} can always handle such corner cases because ``call'' instruction will always change the stack pointer location and  ``jmp'' instruction will not. \Scaler{} can distinguish which API is called by ``jmp'' and which is not by comparing the location of the return address on the stack. \Scaler{}'s ``jmp'' detection mechanism can generalize to the case where there are multiple ``jmp'' instructions to invoke an API. For example: funcA -> call API1@plt -> jmp API2@plt -> jmp API3@plt. In this case, the return address will be somewhere inside funcA, and Scaler should return to funcA. Scaler’s API interceptor will be invoked three times. In every invocation, Scaler’s API interceptor will see that the location of the return address stays the same on the stack. So when Scaler returns to funcA, it will know that API1, API2, and API3 have all finished execution.
}

%% file: src/5_discussion.tex
\section{Discussion \label{sec:Discussion}}

% \ST{Discuss bugs that scaler cannot help with.}
\subsection{Limitation}
\label{sec:limitation}

\SC{} cannot detect hardware-related performance bugs, e.g., cache issues,  especially those ones caused by application code. Such performance bugs can be detected better using \perf{} or CachePerf~\cite{CachePerf}. Currently,   
\SC{} does not intercept internal functions of applications or APIs of statically linked libraries, because these functions or APIs do not use the linkage table. Since \Scaler{} works inside user space, it does not intercept kernel functions.

\SC{} helps identify the number and percentage of API invocations, which share the same purpose as \perf{} or gprof. However,  all three tools still require human effort to actually find out the root cause of performance issues. 

%As mentioned in Section \ref{sec:DesignAndImpl}, a small amount of no-return functions inside libc, such as \_longjmp and \_\_cxa\_throw cannot be intercepted because there is no reliable and efficient way to determine the correct return address in the \texttt{Post-Interceptor}.

% \subsection{Compatibility}
% We implemented the proposed methods \quotes{lightweight PLT interposition}, \quotes{online data folding}, and \quotes{multithreaded runtime attribution} on the x86-64 platform on the Linux system. Nevertheless, the proposed mechanism can be extended to other processor architectures and systems since the dynamic linking mechanism of different systems is similar.

\revision{Supporting static linking libraries is a future research direction, but the proposed techniques such as Universal Shadow Table and Relation-Aware Data Folding can still be applied to profile statically linked libraries. The major technical challenge is to identify statically linked APIs inside compiled programs. But this problem can be bypassed if the symbol table is provided to \Scaler{}. Once the address of a static linking API is known, Scaler needs to replace the first two instructions in the function to make the program jump to the universal shadow table. The instructions used for replacement are similar to what \Scaler{} currently uses to replace the \texttt{.plt}. Similar to simulating instructions inside the \texttt{.plt}, \Scaler{} needs to simulate the behavior of the overridden instructions. This simulation can be achieved by copying the replaced instructions to other memory regions and modifying jump targets and relative addressing instructions accordingly. Since \Scaler{} only overrides two instructions, the simulation should not significantly impact the performance overhead. }

\subsection{Extensibility}
\Scaler{} is easy to extend, and can be connected with different performance analysis tools. For example, it could collect all synchronizations (and parameters) for synchronization analysis~\cite{SyncPerf, wPerf}; the whole sequence of API invocations will provide insights for tail performance analysis~\cite{Dean2013Tail}. 
%\Scaler{} may also be extended to collect certain call stacks based on \XFA{} result so it is easier to find the performance bug. 
Despite these possibilities, \Scaler{} is not designed as the replacement for sampling-based tools (e.g., \perf{}) but as a complement to these existing profilers. 

%% file: src/6_relatedwork.tex
\section{Related work}
 \label{sec:RelatedWork}

\paragraph{General Profilers}
General profilers typically help identify the performance issues of applications, such as  \texttt{gprof}~\cite{gprof}, \-\texttt{Oprofile}~\cite{OProfile}, \-VIProf~\cite{VIProf}, \-\texttt{Coz}~\cite{Coz}, vtune~\cite{vtune-ums}, and \texttt{perf}~\cite{perf}. As discussed in Section~\ref{sec:Introduction}, they typically focus on performance issues of applications but fall short in diagnosing the inefficiency caused by external components. For instance, \texttt{gprof}~\cite{DBLP:conf/sigplan/GrahamKM82} instruments the entry and exit of each internal function with the compiler-assisted instrumentation in order to collect and analyze the execution time of the whole application. However, it explicitly skips the external libraries, as it assumes that libraries are not performance bottlenecks and it is not convenient to recompile all libraries. \texttt{perf}~\cite{perf} cannot precisely diagnose the inefficiency of external libraries due to its coarse-grained sampling. In contrast, \SC{} identifies the inefficiency of external libraries or internal design issues by abnormal API invocations, which could be complementary to existing profilers.

\paragraph{Holistic Profilers}
Holistic profilers focus on providing a holistic view of performance to identify the performance inefficiency in the whole system stack. Stitch~\cite{zhao2016non} profiles the performance of the whole software stack based on the unstructured logs output. Its effectiveness highly relies on the comprehensiveness of logging. In contrast, \Scaler{} does not rely on the program code. Caliper~\cite{boehme2016caliper} is a library-based approach that allows tool developers and users to collect hardware-related events and timestamp information. Caliper requires users to explicitly utilize their provided APIs to collect the data, which is not transparent to users (and therefore different from \Scaler{}). \SC{} complements these profilers with its transparency.
%Further, their performance overhead can be as high as 26\% based on their reported data. \texttt{XSP}~\cite{li2020xsp}, a whole-stack machine learning profiler, does explicitly exclude the external libraries into consideration when profiling the model components, which is significantly different from \Scaler{}. Overall, \Scaler{} could complement the existing holistic profilers due to its unique properties in transparency and efficiency. 
%In the following, we only discuss existing work without changing programs or recompilation, which excludes compiler-based instrumentation. 
%In particular, \texttt{strace} traces system calls and signals, while \ltrace{} can collect and print information about library calls, signals, and system calls. The \texttt{ptrace} system call enables the collection of API invocations without changing the user code. However, 

\paragraph{Library API Interposition:}
There are the following ways of intercepting library API invocations. \textbf{Ptrace-based Approaches: } Ptrace-based tools, such as \texttt{ltrace}~\cite{ltrace} and \texttt{strace}~\cite{strace}, leverage the \texttt{ptrace} system call to intercept functions, without code change. However, typically such tools impose significant performance overhead, e.g., \ltrace{} 's overhead is over $6,100\times$, which may introduce correctness issues for data collection. \textbf{Library Preloading}: The library preloading allows the interposition of library APIs without changing the source code, such as SyncPerf~\cite{SyncPerf} or wPerf~\cite{wPerf}. The preloading mechanism requires users to provide the signatures of interception APIs. \textbf{Changing Dynamic Loader: }  Zaslavskiy et al. ~\cite{zaslavskiy2013lightweight} provided a customized dynamic loader to perform performance profiling. However, this method may introduce compatibility issues when the loader is updated. \textbf{Binary Instrumentation:} Binary instrumentation, e.g., Pin~\cite{Pin} or DynamoRIO~\cite{DynamoRIO},  can interpose library APIs without the explicit change of applications. However, such approaches typically impose more than $5\times$ performance overhead, making the precise time consumption analysis implausible. \textbf{\texttt{.got.plt} Interposition: } \ditool{}~\cite{ditools} proposes to change the addresses stored in the \texttt{.got.plt} so that it can re-direct the flow to the user-defined functions, which also do not need the code change or recompilation. However, \ditool{} requires users to provide the signatures of intercepting APIs, and it strongly couples with the implementation of \texttt{Irix} and \texttt{rld} (a specific version of tools). In contrast, \SC{} overcomes these issues: it can interpose any APIs without the code change or knowing the signatures by employing Selective Binary Instrumentation and Universal Shadow Table.

%supports MIPS architecture and the IRIX system. Second, it does not solve reliability challenges mentioned in Section \ref{sec:DesignAndImpl} such as Eager/Lazy loading; Premature function call; Randomized PLT; Abnormal exits; Irregular API invocations; No-return API support. Third, \ditool{} is not designed for performance profiling and may have high performance overhead due to the lack of crafted assembly instructions and optimizations as mentioned in Section \ref{sec:DesignAndImpl}. In conclusion, \ditool{} cannot satisfy the performance and reliability  challenges as described in Section~\ref{sec:Introduction}; 

%Based on our understanding, some optimization techniques, such as saving API ID inside PLT entry, cannot be applied directly here. Therefore, they cannot work as efficiently as \Scaler{}'s PLT interposition.

%are using the library preloading to collect synchronization events without the need to change programs. However, the preloading mechanism has a very obvious shortcoming: users must provide the complete signatures of APIs, which can be done quickly for synchronization APIs. However, this is implausible for \Scaler{} as it would intercept all library APIs, even when the compiler strips all symbols (including API signatures).  

%% file: src/7_conclusion.tex
\section{Conclusion \label{sec:Conclusion}}

Modern systems are complex, since there exist frequent interactions among all components of the whole system stack. This paper proposes a novel cross-flow analysis  method (called \XFA{}) that helps users understand the interactions of all components. We also implemented a profiler (named \SC{}) that introduces multiple novel techniques that work together to reduce performance and memory/storage overhead, including Universal Shadow Table, and Relation-Aware Data Folding. Our comprehensive experiments confirm that \Scaler{} could identify multiple performance issues that existing tools (e.g., \perf{}) cannot detect. Therefore, \SC{} will be a complement to existing profilers due to its unique property and reasonable overhead. 
%Overall, \Scaler{} only requires around $14.6\%$ performance overhead, which is even lower than sampling-based tools but collects data $772\times$ more frequently. The memory overhead of \Scaler{} is around $10.0\%$, which is orders of magnitude lower than existing tools. 

\section{Acknowledgements}

Many thanks to our shepherd and the anonymous reviewers for their insightful comments. This material is based in part upon work supported by the NSF grant CCF-2118745, CCF-2024253, CNS-2312396, DUE-2215193, and CNS-1750760.